 \documentclass[final,5p,times,twocolumn,authoryear]{elsarticle}

\usepackage{amssymb}
\usepackage{graphicx}
\usepackage{subcaption}
\usepackage{multirow}
\usepackage[ruled]{algorithm2e} 

\usepackage{float}
\usepackage{booktabs}
\usepackage{makecell}  
\usepackage{tabularx}
\usepackage[T1]{fontenc}
\usepackage{booktabs} 
\usepackage{enumitem}

\journal{Journal of Computer and Network Applications}

\begin{document}

\begin{frontmatter}


\author{Baris Yamansavascilar\corref{cor1}\fnref{label1}}
\ead{baris.yamansavascilar@boun.edu.tr}
\author{Ahmet Cihat Baktir\fnref{label1}}
\ead{cihat.baktir@boun.edu.tr}
\author{Atay Ozgovde\fnref{label2}}
\ead{aozgovde@gsu.edu.tr}
\author{Cem Ersoy\fnref{label1}}
\ead{ersoy@boun.edu.tr}


\cortext[cor1]{Corresponding author.}


\title{Fault Tolerance in SDN Data Plane Considering Network and Application Based Metrics}

\address[label1]{Department of Computer Engineering, Bogazici University, Istanbul, Turkey}
\address[label2]{Department of Computer Engineering, Galatasaray University, Istanbul, Turkey}



\begin{abstract}
Failures in networks result in service disruptions which may cause deteriorated Quality of Service (QoS) for the end users. Since SDN is becoming the mainstream paradigm for networks, implementation of a robust fault tolerance scheme for SDN-based networks is crucial. Existing SDN data plane fault tolerance approaches can be classified as reactive and proactive which may or may not rely on the controller, respectively. However, none of them qualifies as a complete solution, providing only partial remedies. In this work, we propose Dynamic Protection with Quality of Alternative Paths (DPQoAP) that considers not only the existing faults within the network but also the quality of alternative paths. As a result, we can sustain the QoS throughout the network after the recovery. We also investigate how application based parameters are affected by link failures. To this end, we explore the change in Quality of Experience (QoE) caused by link failures under different cases using Dynamic Adaptive Streaming over HTTP (DASH) for video streaming. On the other hand, even though DASH is proposed as a solution to improve the QoE affected by the dynamic conditions of the networks, it remains insufficient to handle the congested links that show the symptoms of a link failure. Thus, we apply the data plane fault tolerance approach in SDN to improve the QoE of DASH clients in the case of congestion as well as the failure. The performance of the proposed solutions are evaluated through various experiments considering the QoS and QoE parameters. It is observed that DPQoAP enhances the efficiency of the networking operations and adaptability of the applications. 

\end{abstract}



\begin{keyword}


SDN \sep Fault Tolerance \sep Reliability \sep Dynamic Adaptive Streaming over HTTP (DASH) \sep Quality of Experience (QoE) \sep Quality of Service (QoS)

\end{keyword}

\end{frontmatter}


\section{Introduction}
\label{intro}
Alleviating failures is crucial for service providers in order to meet the Quality of Service (QoS) expected by their clients \citep{gozdecki2003quality}. Strict Service Level Agreement (SLA) requirements when combined with the risk of reputation loss are evident that fault handling should be carefully accomplished. Ideally, network failures should be handled seamlessly, transparent to the end user, not affecting their Quality of Experience (QoE). Practically, however, remedies try to avoid disruptions in the network level QoS and application level QoE as much as possible.

A fault once occurred in a network evidently impacts its operations. This impact can be broadly categorized as the impact on QoS behavior and the effect on QoE behavior of the network. Here, the QoS parameters represent the application independent network characteristics whereas QoE parameters represent the specific impact that an end user experiences for a given application. Both QoS and QoE parameters are of importance for assessing how a given fault triggers fluctuations in the network performance.

In the context of fault tolerance, Software-Defined Networking (SDN) provides important opportunities with its central view on the whole network. In SDN, the concept of fault tolerance can be taken into consideration within three domains: the data plane, the control plane, and the application plane. The data plane issues consist of link or switch failures whereas the control plane domain considers the failure of the switch controller connections or the failure of the controller itself. The application plane domain focuses on the failure of an application that can affect the northbound API which in turn can gradually affect other applications. In this study, we focus on the fault tolerance within SDN data plane.

\begin{figure}[t]
\centering
\includegraphics[scale=0.37]{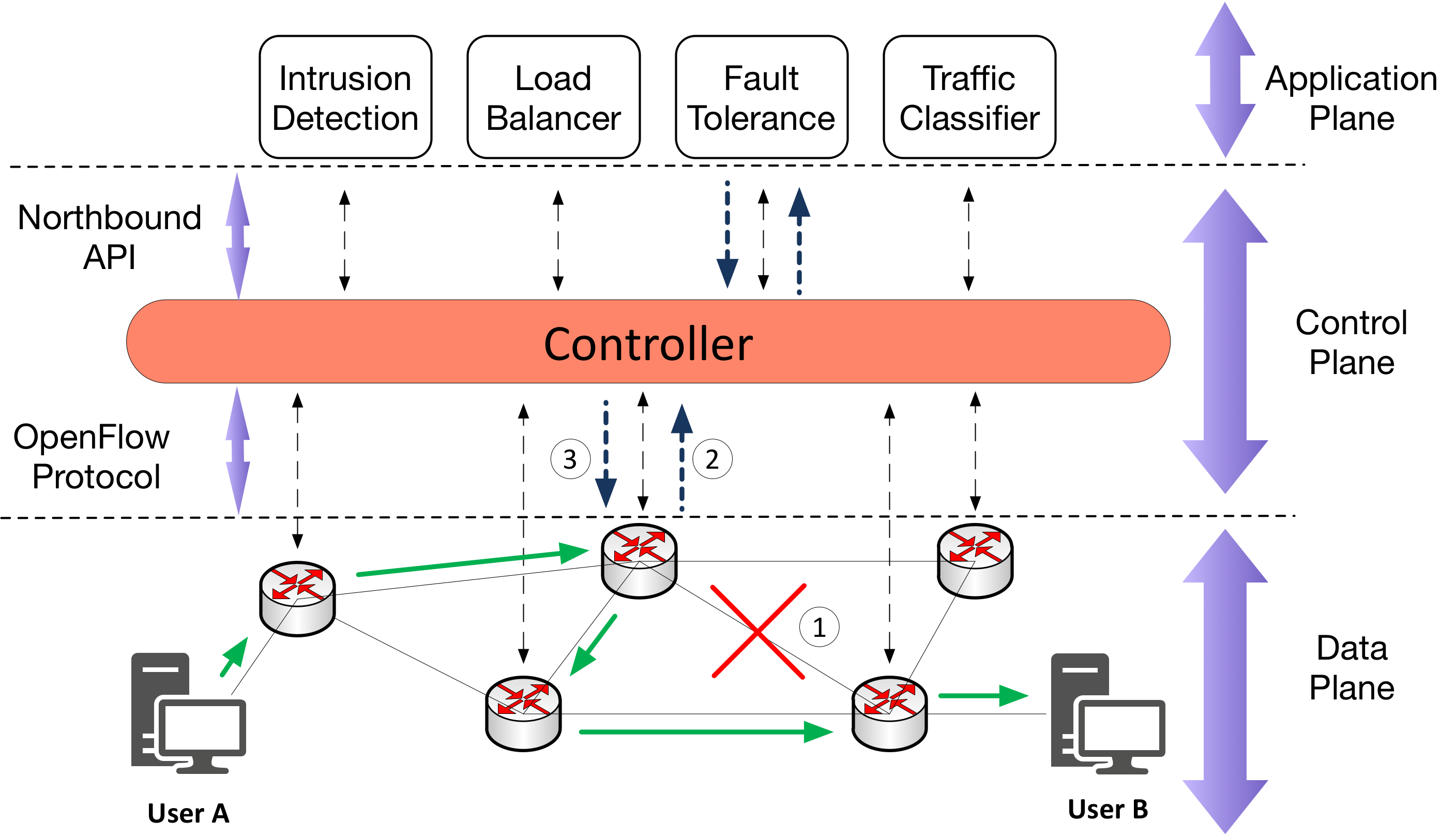}
 \DeclareGraphicsExtensions.
\caption{Restoration in SDN.}
\label{restoration}
\end{figure}

Restoration and protection are two essential approaches for the failure recovery in the data plane. Both of these approaches have their respective advantages and disadvantages in terms of the recovery time and recent network view \citep{fonseca2017survey, yu2018fault}. In restoration, the new routing rules for the affected flows are computed considering the recent network view. During this process, the controller itself is the responsible entity as shown in Figure \ref{restoration}. When a failure occurs in the data plane, it is initially detected by the corresponding switch and then an event is generated for the controller to trigger the path calculation process. Afterwards, the event is processed in the northbound application and the new route is computed. In protection on the other hand, the alternative rules for active flows are predefined for possible failures in the future. Moreover, thanks to the Fast Failover groups proposed in OpenFlow version 1.1, the controller involvement is not necessary to update the flow rules in case of a failure. Thus, the failure recovery time is shorter than the restoration in this approach.

Protection is usually accepted as a superior approach since it reduces the recovery time significantly compared to the restoration. However, this view does not correctly reflect all the requirements of a fault protection mechanism since it omits the quality of the new path substituting the faulty one. Typically, in a network, various alternative paths exist to recover a given fault and the quality of these paths should be incorporated into the selection process. With this vision in mind, we propose Dynamic Protection with Quality of Alternative Path (DPQoAP) which combines both the restoration and the protection methods to achieve fast recovery while selecting among available high-quality paths. To achieve this goal, DPQoAP takes two important requirements into account: (1) the recent state of the network considering restoration, and (2) backup path information for the active flows regarding protection. To carry out first requirement, DPQoAP periodically checks the quality of alternative paths in the network. For the second requirement, it uses Fast Failover groups to hold backup path information in corresponding switches. 

To evaluate the performance of DPQoAP, we focused on the video streaming use case. Currently, video streaming is the most influential traffic type of the Internet since it occupies 73\% of the overall data traffic volume \citep{index2017zettabyte}. Apart from its major role in the traffic composition, video streaming also forms a separate category with respect to other applications since it is continuously evaluated by the end user. Dynamic Adaptive Streaming over HTTP (DASH) is the widely accepted technology recently for video streaming. DASH, can receive video segments in independent connections, which in turn allows for versatile and dynamic streaming operation.

Assuming a continuum of network performance degradation, link failure can be considered as an extreme case of congestion in which the delay becomes infinity. Bearing this view in mind, fault tolerance solutions originally developed for link failures can, in fact, serve as candidate methods to deal with congestion when carefully adjusted to this new context. Accordingly, even though the adaptive nature of DASH provides important flexibility considering the unstable network conditions, its capabilities are insufficient to handle the extreme congestion case as well as the link failure. Thus, we combine the capabilities of DASH and our novel approach that perceives the congestion as the fundamental element for the link failure. Our experimental results show that QoE parameters including video quality, bitrate latency, and the number of quality switches are improved dramatically when we apply our method to the congestion case.  


Another focus of our work is related to the inadequacy of an experimentation method that is typically being used in the literature. Mininet, which owes its reputation to the wide variety of its capabilities, is the emulation environment for evaluating SDN-based proposals \citep{de2014using}. However, in the context of generating link failures, we argue that capabilities provided by Mininet are insufficient to emulate a fault scenario in a realistic manner since it actually destroys the whole connection including the ports of switches. Moreover, Mininet allows switches to notify port failures to the northbound applications with almost zero delay which can cause incorrect recovery times to be reported. As a remedy, in our work, we incorporate L2 Linux bridges \citep{varis2012anatomy} that are transparent to both the controller and the switches to design the failure experiment scenarios. We demonstrate that our approach more realistically emulates network failures even in Mininet.

In this study, we focus on the change of network and application based metrics regarding QoS and QoE, respectively, in case of a failure in the data plane. Moreover, we investigate the effect of congestion on QoE. Thus, main contributions of our study can be summarized as:

\begin{itemize}

\item We propose DPQoAP which is a novel dynamic protection method considering the quality of alternative paths. In this case, in addition to the recovery time, we also consider the alternative paths for flows affected by the failures. 

\item We investigate how video quality and thus QoE would fluctuate when a failure occurs. Since related studies in the literature consider only the variations of network parameters regarding QoS, this study carries out a unique touch to this problem considering the change of application parameters.

\item Studies using an emulation environment such as Mininet for SDN do not consider the broken link scenario that may emulate the real-world case and therefore trigger the failure by using a special Mininet command. In this study, to emulate this realistically, we deploy a Linux bridge that operates in L2, and is completely transparent to the controller and switches in the network for particular scenarios in order to make a valid comparison and evaluation.

\item We detect the congestion in the link layer operating Bidirectional Forwarding Detection (BFD) protocol rather than the application layer metrics that is generally used by video applications to adapt themselves for new conditions. Detecting congestion at the network level through SDN allows us to implement a global solution as opposed to each client adapting themselves in a reactive manner.

\item After its successful detection, we employ the data plane fault tolerance mechanism to solve the congestion problem for DASH in SDN environments. As a result, we apply the same treatment carrying out for the link failure, to a different problem, congestion, that shows similar symptoms.

\item We explore the impact of BFD intervals on the QoE parameters along with the video segment size to come up with a feasible operational range where the fault tolerance approach is beneficial for the congestion case.

\end{itemize}

The rest of this study is organized as follows. In Section II, we elaborate the related works including DASH, fault tolerance, and congestion studies that use SDN as their networking paradigm. Section III provides details of our fault tolerant design. Afterwards, in Section IV and V, we present the performance evaluation of our system along with the experiments. Finally, we conclude this study in Section VI.

\section{Related Works}
\label{related}

SDN, with its dramatically different paradigm, offer many benefits to address networking challenges that are hard to circumvent with traditional approaches. The existence of a centralized controller in addition to the radically different flow of events provided by SDN render these solutions technically feasible. One such challenge is the fault tolerance including data plane, control plane, and application plane domains. Fault tolerance in each domain leads to different problem formulations. This study focuses on the data plane domain as it is the level where the actual traffic is handled.

\subsection{Restoration Approaches}

Considering fault tolerance in SDN, one of the most important concepts in OpenFlow protocol is the Fast Failover (FF) groups that allow for solving faults in the data plane without involving the controller. However, before OpenFlow protocol version 1.1 in which OpenFlow Groups including Fast Failover groups are introduced, many studies \citep{sharma2011enabling, kim2012coronet, nguyen2013software, li2014scalable} that work on this topic used the restoration approach. The common behavior in these studies is that they used the controller for failure notification and then calculate new routes to maintain communication. Kim et al. \citep{kim2012coronet} used VLANs to compute routing paths. In \citep{sharma2011enabling}, Sharma et al. compared their fast failover system with the Learning Switch, Learning PySwitch, and Routing Mode of the NOX controller \citep{gude2008nox}. Nguyen et al. \citep{nguyen2013software} used SDN for fault tolerance in Wide Area Networks (WANs) since routing protocols such as BGP and OSPF undergo serious problems in failures. For example, while BGP has prolonged route convergence time, OSPF has long recovery time. Li et al. \citep{li2014scalable} on the other hand developed a restoration approach by using a local optimal failover method in order to reduce the path calculation time. Some studies also considered reliability in this manner. In \citep{yuan2018practical} Yuan et al. designed a system based on the Byzantine model to tolerate faulty switches in order to enhance reliability. Moreover, Song et al. \citep{song2017control} focused on control-path reliability which is an important consideration for out-of-band controllers whose network view can be affected by the data plane failures.

\subsection{Protection Approaches}

Since using the controller for fault tolerance increases the recovery time, many studies proposed methods to solve this problem in the data plane without involving the controller. In \citep{desai2010coping}, authors built a system that permits switches to send faulty link information to only the relevant switches in order to prevent traffic flood and increase the network performance in centralized networks. Thus, when a link fails in the network, the corresponding switches create Link Failure Messages (LFM) and send them to the relevant switches. Their experiments showed that switches are informed of the failed link sooner than the controller identifies and commits an update. Likewise, Kempf et al. \citep{kempf2012scalable} supported a monitoring function for failure detection in the data plane without involving the controller. To this end, they generated monitoring messages within the source switch and process them in another destination switch. If the destination cannot receive these packets for a long enough period, their method concludes that there is a fault in the current path. Their experiments showed that the data plane fault recovery can be achieved in a scalable way within 50 milliseconds using this function. In \citep{ramos2013slickflow}, Ramos et al. extended their previous study \citep{ramos2013data} and developed a proactive failure recovery scheme by carrying the information of alternative paths in the packet headers. Thus, when a link failure happens, their system uses alternative path information in order to maintain communication without consulting the controller. In the packet header, they used VLAN and MAC Ethernet fields to carry alternative paths.

On the other hand, Reitblatt et al. \citep{reitblatt2013fattire} proposed a new language based on Regular Expressions for implementing fault-tolerant network programs in SDN by using OpenFlow Fast Failover groups. They allowed developers to specify the set of paths that packets may take through the network as well as the degree of fault tolerance required. Thus, their compiler generated rule-tables and group-tables that provide specified fault tolerance. Accordingly, Petroulakis et al. \citep{petroulakis2017fault} proposed a pattern framework for fault tolerance using a rule-based language. Moreover, Cascone et al. \citep{cascone2017fast} used finite state machines in the data plane for fast failure detection and then recovery.

After OpenFlow protocol version 1.1 was introduced, studies used Fast Failover groups as it provides many benefits in fault tolerance including recovery time and control-path traffic. To this end, Sharma et al. \citep {sharma2013openflow} considered carrier-grade networks in which fault recovery should be completed in 50ms, and therefore they performed protection mechanisms using OpenFlow Fast Failover groups. The experimental results showed that the protection approach diminishes the time required for fault recovery and mitigates the traffic load on the controller. Moreover, in \citep{sharma2013fast}, they focused on failure recovery for the in-band OpenFlow networks in which control and data traffic are transmitted on the same channel by applying the same scenarios. In \citep{borokhovich2014provable}, Borokhovich et al. implemented traditional graph algorithms including BFS, DFS, and Module to compute backup paths which are used in FF groups. Adrichem et al. \citep{van2014fast} used Bidirectional Forwarding Detection (BFD) protocol \citep{rfc5880} per link as well as \citep{sharma2013openflow} to detect failures and then compared the performance of different BFD detection intervals in terms of milliseconds. Pfeiffenberger et al. \citep{pfeiffenberger2015reliable} on the other hand, focused on the robust multicasting in SDN by considering fault tolerance. In \citep{thorat2017rapid}, the authors used VLAN tags in their design in order to reduce alternative path rules.

\subsection{DASH and QoE in SDN}

Recently, studies working on DASH use SDN as the main networking technology in order to exploit the benefits of the centralized controller and, thus, enhance QoE. Zabrovskiy et al. \citep{zabrovskiy2016emulation} applied DASH in Mininet and compare its performance with a specialized hardware-software emulator using the same channel characteristics. The performance results obtained from both settings were comparable, which enabled authors to conclude that Mininet can be used as a reliable emulator for DASH. Mkwawa et al. \citep{mkwawa2016video} proposed a video quality management scheme that considers the traffic intensity. They compared the number of stalls of the DASH video streaming when their proposed scheme is used and not used. As expected, the number of stalls were fewer when they used their solution. In \citep{bentaleb2016sdndash}, Bentaleb et al. focused on QoE unfairness for multiple clients in the network since when the number of clients increases, QoE is unfair because of bandwidth sharing and network resource underutilization. Afterwards, they improved their scheme in \citep{bentaleb2017sdnhas}, considering scalability issues of clients, communication overhead, and client heterogeneity. Likewise, Bagci et al. \citep{bagci2017compete} studied QoE fairness among clients. They used the network slicing concept and manipulated TCP windows to prevent QoE fluctuations.

\subsection{Congestion in SDN}
SDN is also used for the congestion case by several studies. In \citep{lu2015sdn}, the authors modified the TCP receive window of ACK packets at the controller in order to avoid network congestion. To perform this, they deployed a queue management scheme in OpenFlow-switch that notifies the controller when the queue passes the given threshold. Kim et al. \citep{kim2016congestion} on the other hand considered the dynamic changes in the network traffic and proposed their reinforcement learning based technique, Q-learning, for the routing of flows to prevent congestion. Cheng et al. \citep{cheng2017congestion} indicated the shortage of ternary content addressable memory (TCAM) of OpenFlow switches that cause a bottleneck for scalable flow management. Therefore, they applied flow aggregation using VLANs to prevent congestion for a failure recovery case. In \citep{nasimi2018edge}, the authors carried out a congestion control mechanism in Mobile Edge Computing (MEC) environment using SDN considering congestion. They classified the network traffic as delay tolerant and delay sensitive so that they buffered the delay tolerant flows in MEC servers during the peak hours in order to prevent congestion.

To the best of our knowledge, there is no study in the literature that examines QoE for fault tolerance in SDN considering video streaming using the DASH paradigm. Moreover, studies worked on fault tolerance considered only QoS parameters including recovery time, delay, and the number of affected flows rather than application parameters that affect QoE. On the other hand, this is the first study that applies a fault tolerance approach to improve QoE in case a network-based problem like congestion. Thus, our study is distinctly separated from the existing works.



\section{Design of the Fault Tolerant Data Plane}

The static protection has many benefits over the restoration as the failure is handled in the data plane. However, the performed action using the Fast Failover groups for the failure is actually based on the network information that is taken at the beginning of the communication. Therefore, since the network environment can change over time because of various reasons, the applied actions may not be efficient for the communication even though it provides the continuity of it. Thus, in DPQoAP, we combine the advantages of the restoration module in which the actions are applied using the recent condition of the network, and the static protection module which provides responsiveness for the failures without involving the controller.

\subsection{OpenFlow Groups}

\begin{figure}[t]
\begin{subfigure}{0.23\textwidth}
\includegraphics[width=\linewidth]{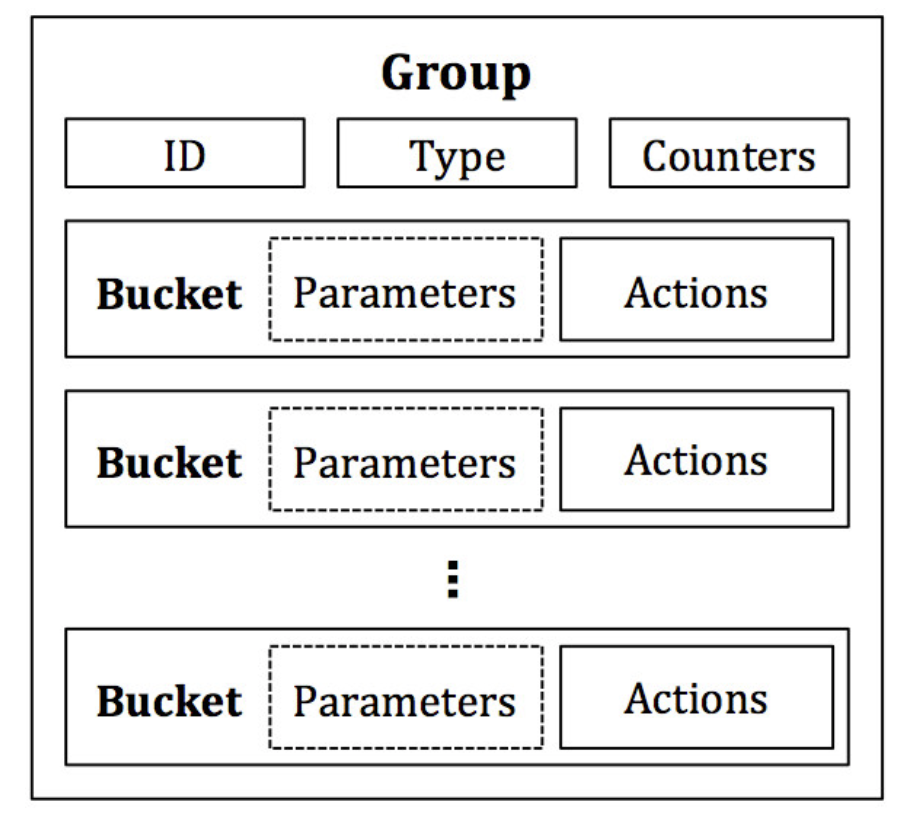}
\caption{Standard OpenFlow group.} \label{StandardOpenFlowGroup}
\end{subfigure}
\hspace*{\fill} 
\begin{subfigure}{0.23\textwidth}
\includegraphics[width=\linewidth]{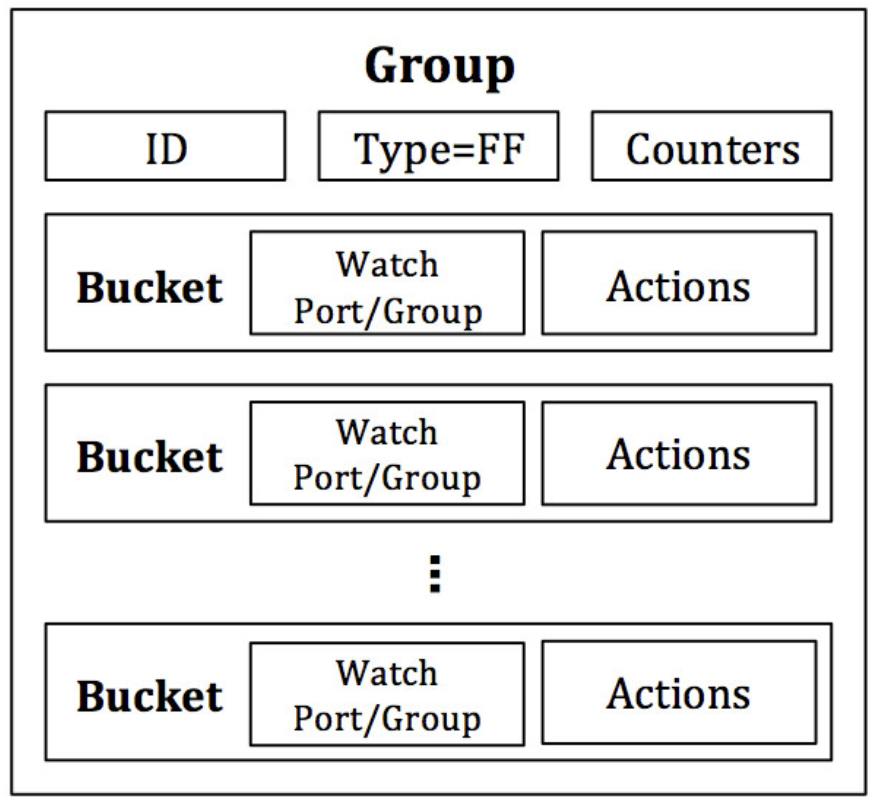}
\caption{OpenFlow fast failover group.} \label{FastFailoverGroup}
\end{subfigure}
\caption{OpenFlow Groups.} \label{GeneralOFGroup}
\end{figure}

In the first stable version of OpenFlow, namely 1.0, there was no specific functionality for fault tolerance. Therefore, network managers and researchers used their own techniques to overcome the failures in SDN until OpenFlow version 1.1 in which the group table concept was introduced. The goal of the OpenFlow groups is to apply specific operations that cannot be defined by the flow itself such as backup path information on packets. A group consists of entries that have a group identifier, a group type, counters, and a list of action buckets as shown in Figure \ref{StandardOpenFlowGroup}. One of the most important features brought by OpenFlow groups is the ability to define multiple lists of actions, which is called an action bucket, for each group entry. This feature makes possible to perform various traffic engineering operations in SDN. When a packet matches with a flow rule, it is assigned to the appropriate group entry. There are currently four OpenFlow group types:

\begin{itemize}
  \item \textbf{All Group:} This group is used for multicasting or broadcasting. A packet in this group is copied for each bucket in the bucket list and handled independently.
  \item \textbf{Select Group:} It is developed for load balancing. The packet in the group entry is sent to a single bucket. Determining the corresponding bucket is performed by the switch itself with a selection algorithm such as round robin.
  \item \textbf{Indirect Group:} There is only one bucket in this group type. The goal is to cover commonly used actions for the same next hop when forwarding and thus reduce the switch memory utilization.
  \item \textbf{Fast Failover Group:} Likewise, in the select group, the packet is sent to only a single bucket. The difference is that there is no bucket selection in this group. The packet is sent to the first live bucket. The liveness of bucket is checked by \textit{watch port/group} parameters. The schema of this group is depicted in Figure \ref{FastFailoverGroup}.
\end{itemize}

\begin{figure}[t]
\centering
\includegraphics[scale=0.10]{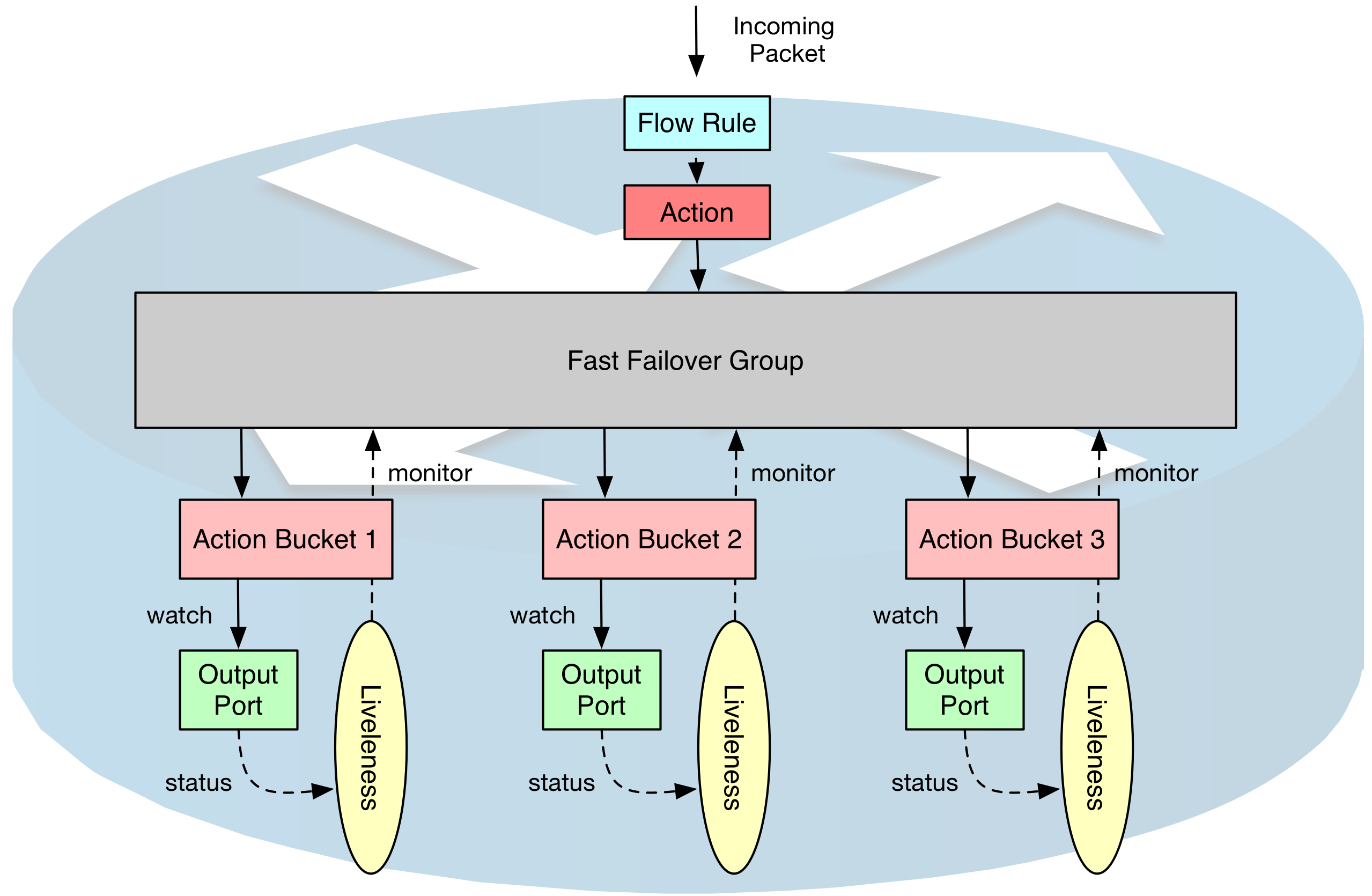}
 \DeclareGraphicsExtensions.
\caption{The working mechanism of a Fast Failover group in OpenFlow switch.}
\label{FFGroupProcess}
\end{figure}

Fast Failover groups are used for the protection approach in SDN. They provide alleviated control path traffic since the controller is not involved in the failure, and reduced recovery time as the problem is solved in the data plane. The working process of Fast Failover group concept is depicted in Figure \ref{FFGroupProcess}. The incoming packet is first evaluated by the flow-rule table and then the corresponding action, which is Fast Failover group in this case, is applied. Afterwards, based on the liveliness status of the possible output ports that are monitored continuously, the packet is sent to the first available egress port. Thus, if a link fails, the appropriate action is instantly applied by the switch without consulting the controller.

\subsection{Bidirectional Forwarding Detection (BFD)}

Among the failure detection methods, the BFD protocol is special since it is designed specifically for failure detection and its detection speed outperforms the others \citep{sharma2013openflow, Adrichem2014}. BFD protocol can be run in computer networks with any transport protocol for fault tolerance since it is protocol-independent. BFD is used for paths between two nodes in order to observe failures and disruptions in communication quickly. These two nodes can be connected either directly or through multiple hops. Each node transmits a control packet including the current state of the monitored link or path to its pair node. When a node receives the control message, it sends an echo message with the session status. Accordingly, failure detection time, $T_d$, is computed based on the message transmit interval, $T_i$, and the detection time multiplier, $M$, as given in Equation \ref{eq2}. The detection time multiplier is used to prevent false positives that can occur due to packet loss. 

\begin{equation}
\label{eq2}
T_d = (M + 1) * T_i
\end{equation}

\subsection{DPQoAP Module}

\begin{figure*}[t]
\begin{subfigure}{0.30\textwidth}
\includegraphics[width=\linewidth]{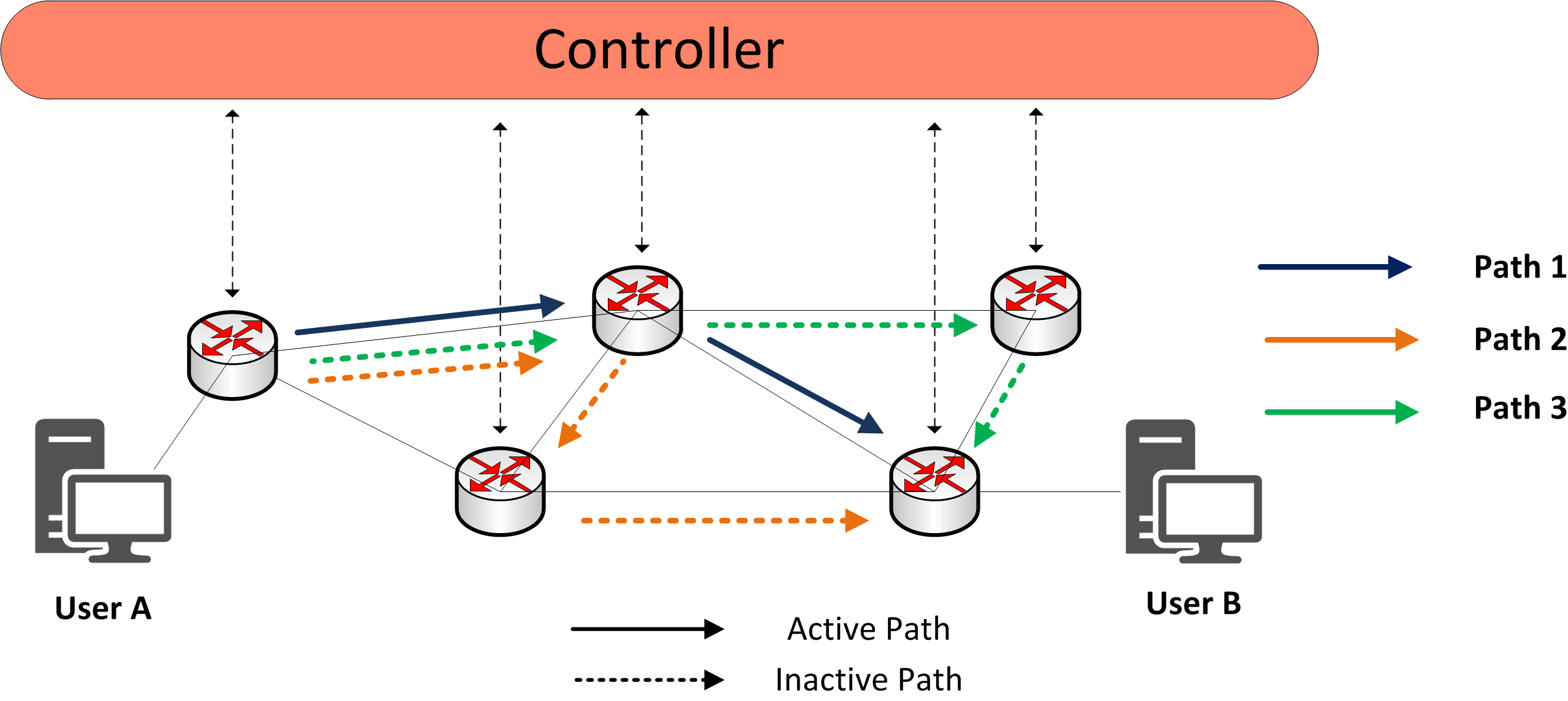}
\caption{Primary and backup paths between User A and User B.} \label{QoAP-A}
\end{subfigure}
\hspace*{\fill} 
\begin{subfigure}{0.30\textwidth}
\includegraphics[width=\linewidth]{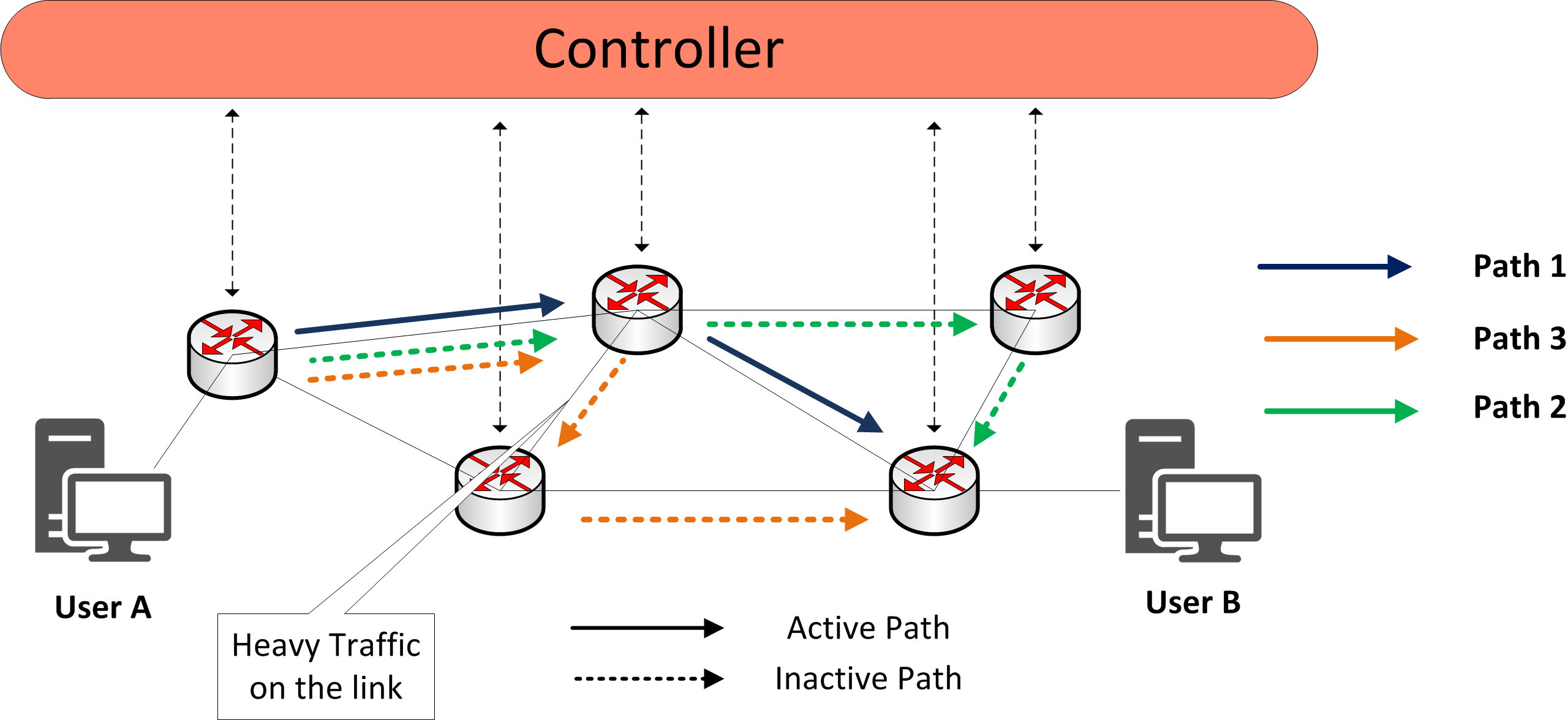}
\caption{Heavy traffic on the original Path 2 and changing the order of the backup paths.} \label{QoAP-B}
\end{subfigure}
\hspace*{\fill} 
\begin{subfigure}{0.30\textwidth}
\includegraphics[width=\linewidth]{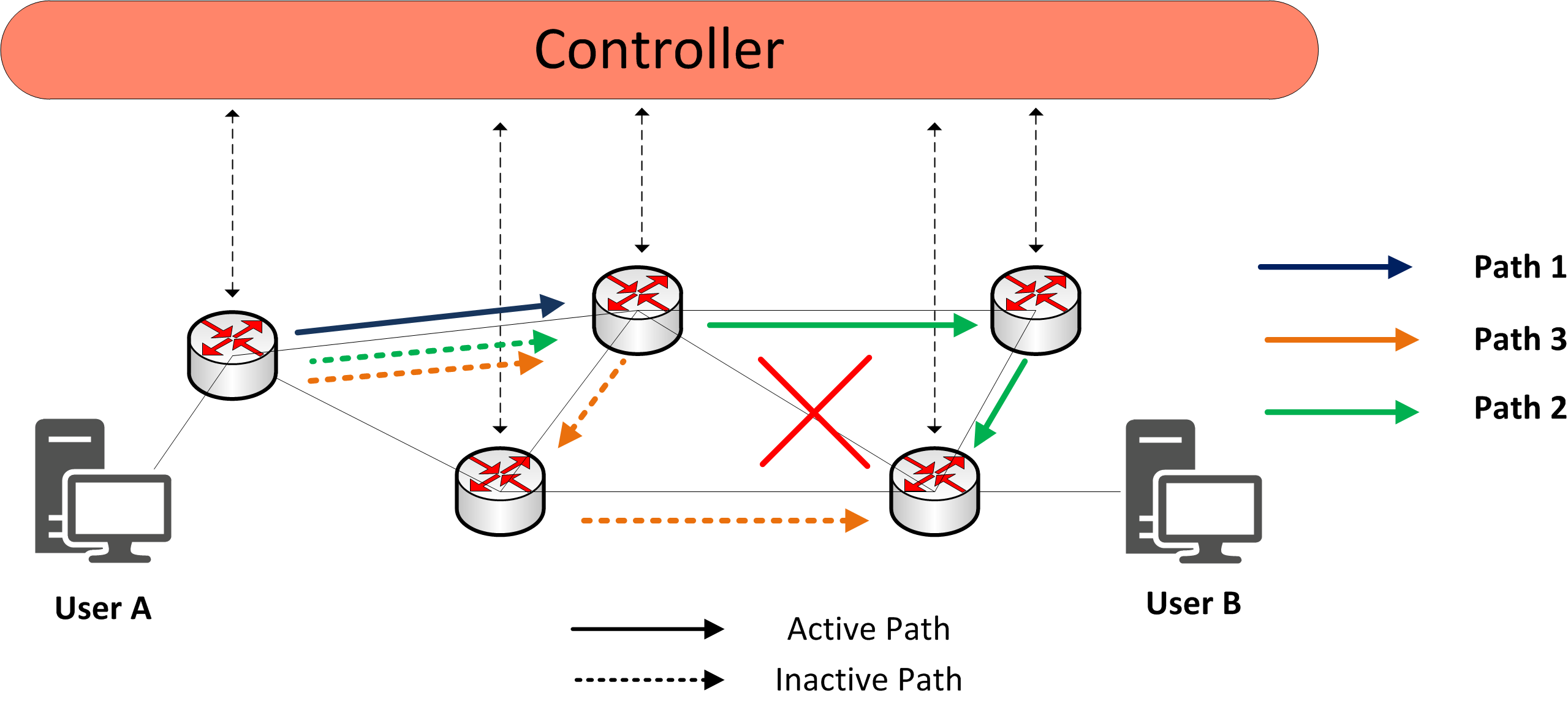}
\caption{Evading the failure in the network considering the QoAPs.} \label{QoAP-C}
\end{subfigure}
\caption{The concept of the dynamic protection module.} \label{ConceptOfDynamicProtection}

\end{figure*}

The main task of this module is to find the best alternative path based on the latency parameter. Since Floodlight controller \citep{floodlight18} provides an API call to obtain the latency values of the links, this information is used for each path to determine the best alternative path for every QoAP calculation interval, {$T_{qoap}$}. A demonstrative example is shown in Figure \ref{ConceptOfDynamicProtection}. After the formation of the primary path and two backup paths as depicted in Figure \ref{QoAP-A}, the secondary path, Path 2, becomes loaded with a recently generated heavy traffic as shown in Figure \ref{QoAP-B}. This situation is detected by our DPQoAP module and subsequently, Path 3 is replaced as the secondary path since its latency value is currently smaller than Path 2. Finally, as shown in Figure \ref{QoAP-C}, when the link of the primary path fails, the quality of the alternative path has already been considered and therefore communication continues with acceptable quality.

We implemented this module as given in Algorithm \ref{dynamicAlgo}. First, the current network state is analyzed in order to get the recent information of the links, switches, and their load conditions. Afterwards, for each $groupID$, which indicates a \textless source, destination\textgreater{} tuple, all computed paths are examined and then the primary path is extracted by checking whether the path is currently active or not. Afterwards, each primary path and its corresponding $groupID$ is given as the parameters to the function $OrganizeBucketList$ in which the best alternative path is calculated from each \textless switch, port\textgreater{} tuple that is in the primary path and then bucket lists are reorganized. This operation is repeated based on the QoAP calculation interval, {$T_{qoap}$}, that must be configured for the network conditions.

\begin{algorithm}
\KwIn{Paths of groups, bucket list for groupIDs}
\KwOut{The Best backup path based on latency}
\SetKw{KwBy}{by}
\SetKwProg{Fn}{Function}{}{end}
\Fn{EvaluationOfPaths ()}{
ComputeCurrentNetworkState()\;
\For{$groupID \in groupIDs$}{
    $groupPaths$ = $pathsOfGroups$($groupID$)\;
     \If{$groupPaths \neq null$}{
     	\For{$path \in groupPaths$}{
		\If{$IsPathActive(path) = true$}{
			$primaryPath$ = $path$\;
			break\;
		}
	}
	OrganizeBucketList($groupID$, $primaryPath$)\;
     }
}
	
}

\caption{DPQoAP Module}
\label{dynamicAlgo}
\end{algorithm}

\subsection{DASH Module}

To calculate the necessary metrics including the buffer level and the bitrate for the video quality value, we implemented a video client module using DASH.js \citep{dashjs}. It is a widely used Javascript library to measure the DASH client metrics. Thus, this module consists of several functions for observing the changes in the metrics in the case of a failure and congestion. 

Apart from the failure, for the congestion scenario that includes multiple clients, each DASH client reports their QoE parameters including the bitrate, video quality value, latency, and the number of quality switches for the evaluation of the effect of congestion. All these parameters except the video quality value are extracted from the DASH.js API.

\subsubsection{Video Streaming and Quality of Experience}

Video streaming has dramatically changed for the last few years due to the recent developments of the Internet technologies and protocols. Considering the increasing usage of mobile devices and the high-resolution options, this change is inevitable. In traditional video streaming, protocols like Real-Time Streaming Protocol (RTSP) behave in a stateful manner via tracking the state of the clients during the streaming. Moreover, when a streaming process has been started between a client and a server, the connection is maintained as the stream of packets until the video file is completely transferred. However, this approach is not sufficient today considering the dynamic needs of video streaming. 

On the other hand, the Hyper-Text Transfer Protocol (HTTP) is stateless; when the client receives its requested data, the connection is terminated. Thus, the streaming is performed in a more dynamic manner when HTTP is used since each HTTP request results in a new transaction that provides many advantages for video streaming \citep{stockhammer2011dynamic}. As a result, Dynamic Adaptive Streaming over HTTP (DASH) is proposed by addressing the weaknesses of the traditional streaming methods.

\begin{figure}

\includegraphics[scale=0.09]{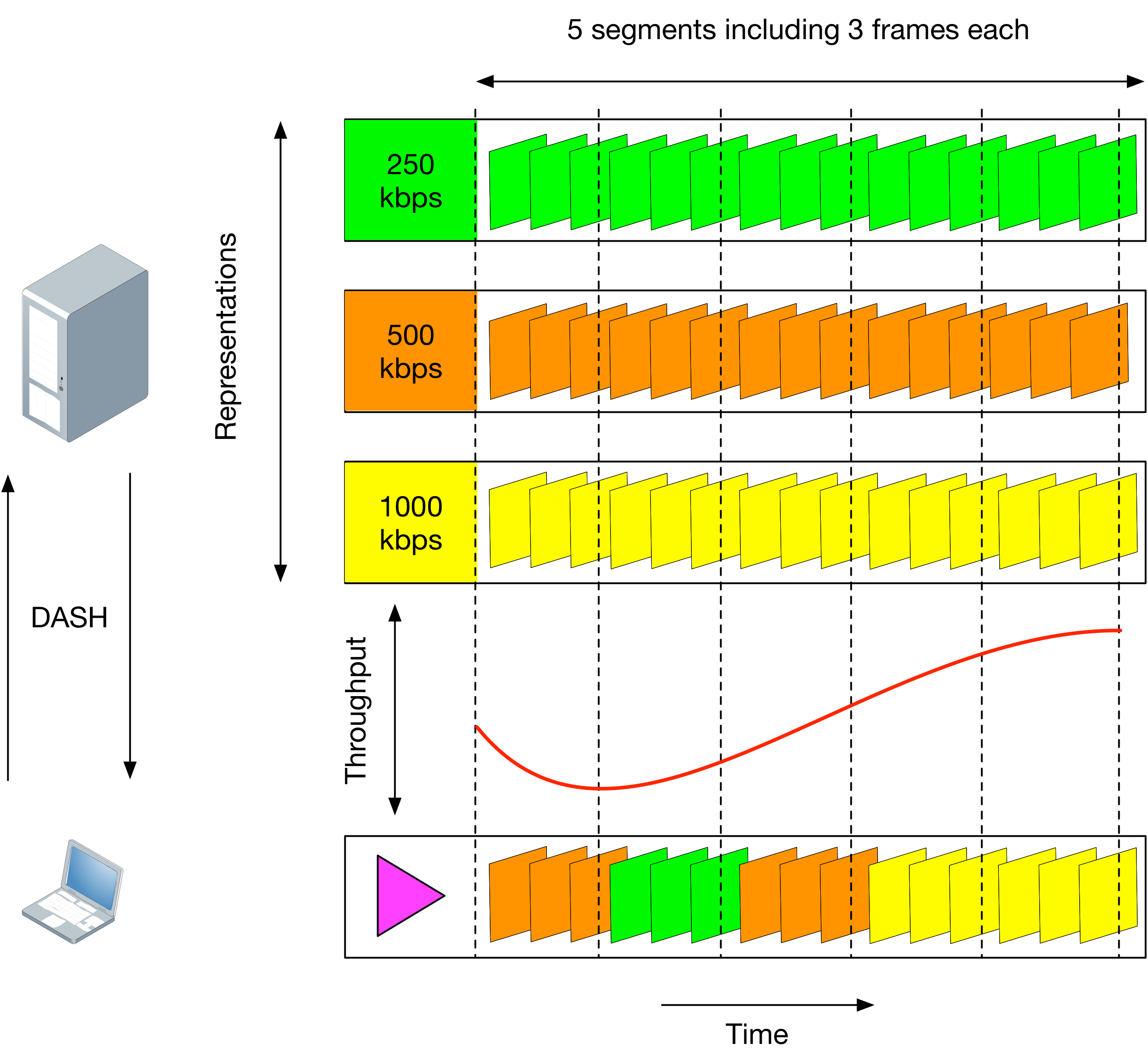} 
\centering
 \DeclareGraphicsExtensions.
\caption{Concept of DASH (Adopted from \citep{seufert2015survey}).}
\label{DASH}
\end{figure}

DASH operates upon fixed durations of video segments called "representations" which may belong to different bitrates. DASH introduces further adaptivity to the dynamic nature of HTTP streaming by enabling a switching mechanism among
different representations. To perform this concept, first, the video file is sliced into the fixed timed parts, namely segments or chunks, at the server as described in a Media Presentation Description (MPD) file. These video parts generally vary from 1 second to 15 seconds of duration. The client can select appropriate segments among different representations based on its application metrics including the buffer level and throughput. Thus, a playback can typically consist of different representation segments instead of homogeneously defined video streaming files. This concept of DASH is depicted in Figure \ref{DASH}.

The most important advantage of DASH considering QoE is the switching mechanism among different representations since it prevents stalling and thus the client can continue to play the video under several conditions. To measure QoE, there are subjective and objective measurements. In subjective measurements, users vote their perception of video using the 5-point scale where 1 represents "poor" and 5 represents "excellent". This process is named as Mean Opinion Score (MOS). On the other hand, in objective measurements, several QoE metrics for DASH \citep{oyman2012quality} including HTTP transactions, representation switch counts, buffer level, and bitrate are evaluated and finally, a formula is generated. Actually, since the video quality is the most distinctive component of QoE, many studies used the video quality as the objective measurement for QoE.

\begin{figure}[!t]

\includegraphics[scale=0.35]{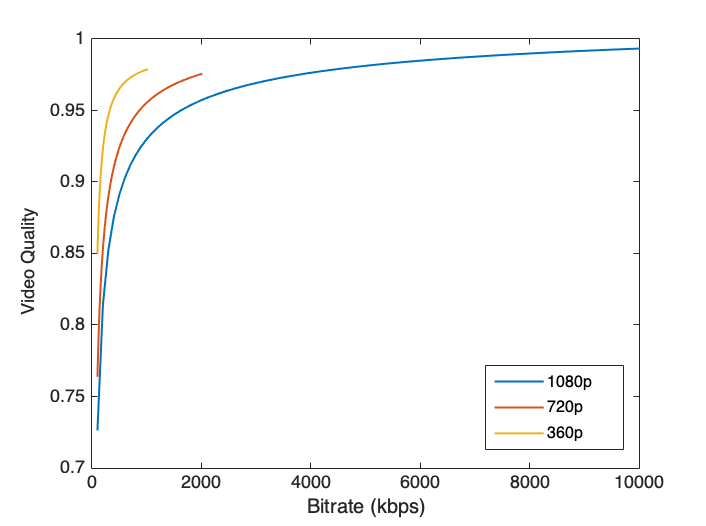}
 \DeclareGraphicsExtensions.
 \centering
\caption{Relation between video quality and bitrate for three different resolutions.}
\label{SSIM}
\end{figure}

One of the most used metrics to measure the quality of the video is the Structural Similarity Index (SSIM) \citep{wang2004image}. It is originally used to measure the quality of an image by comparing it with the original image. Since a video consists of multiple images, the same technique is widely used for measuring the video quality. Thus, mapping the video quality and bitrate is coherent because higher bitrate provides higher similarity with the original video. However, as shown in Figure \ref{SSIM}, the mapping of the bitrate to the video quality using SSIM demonstrated that the relationship between bitrate and perceptual quality is not linear \citep{georgopoulos2013towards}. Hence, using three resolution types and various bitrates given in Table \ref{table1}, a generalized function for QoE based on bitrate and resolution is configured by conducting curve fitting \citep{georgopoulos2013towards}. The formula is given in Equation \ref{eq5} and corresponding coefficients are given in Table \ref{table2}. In the equation, \textit{f(x)} represents the video quality, and variable \textit{x} denotes the bitrate.

\begin{equation}
\label{eq5}
f(x) = ax^b + c
\end{equation}

\begin{table}[b]
\caption{Bitrates for different screen resolutions}
\label{table1}
\centering
\resizebox{\columnwidth}{!}{
\begin{tabular}{| l | l |}
\hline
\textbf{Resolution} & \textbf{Bitrate (kbps)}\\
\hline
1080p & 100, 200, 600, 1000, 2000, 4000, 6000, 8000\\
720p & 100, 200, 400, 600, 800, 1000, 1500, 2000\\
360p & 100, 200, 400, 600, 800, 1000\\
\hline
\end{tabular}
}
\end{table}

\begin{table}[t]
\caption{Coefficients of the generalized video quality function}
\label{table2}
\centering

\resizebox{\columnwidth}{!}{
\begin{tabular}{|c | l| l| l| l | l|}

\hline
 \multirow{2}{*}{\textbf{Resolution}} & \multicolumn{3}{|c|}{\textbf{Power Series Model}} & \multicolumn{2}{|c|}{\textbf{Goodness of Fit}}\\ 
 & \multicolumn{1}{|c}a & \multicolumn{1}{c}b & \multicolumn{1}{c|}c 
& 
\multicolumn{1}{|c}{Adjusted $R^2$} & \multicolumn{1}{c|}{RMSE}
\\
\hline

1080p & 
 -3.035 & -0.5061 & 1.022 
& 
0.9959 & 0.006011
\\
\hline

720p & 
 -4.85 & -0.647 & 1.011
& 
0.9983 & 0.002923
\\
\hline

360p &
 -17.53 & -1.048 & 0.9912
&
 0.9982 & 0.002097
\\
\hline

\end{tabular}
}
\end{table}

\subsection{BFD-based Congestion Detection Module}

\begin{figure}[t]
\centering
\includegraphics[scale=0.085]{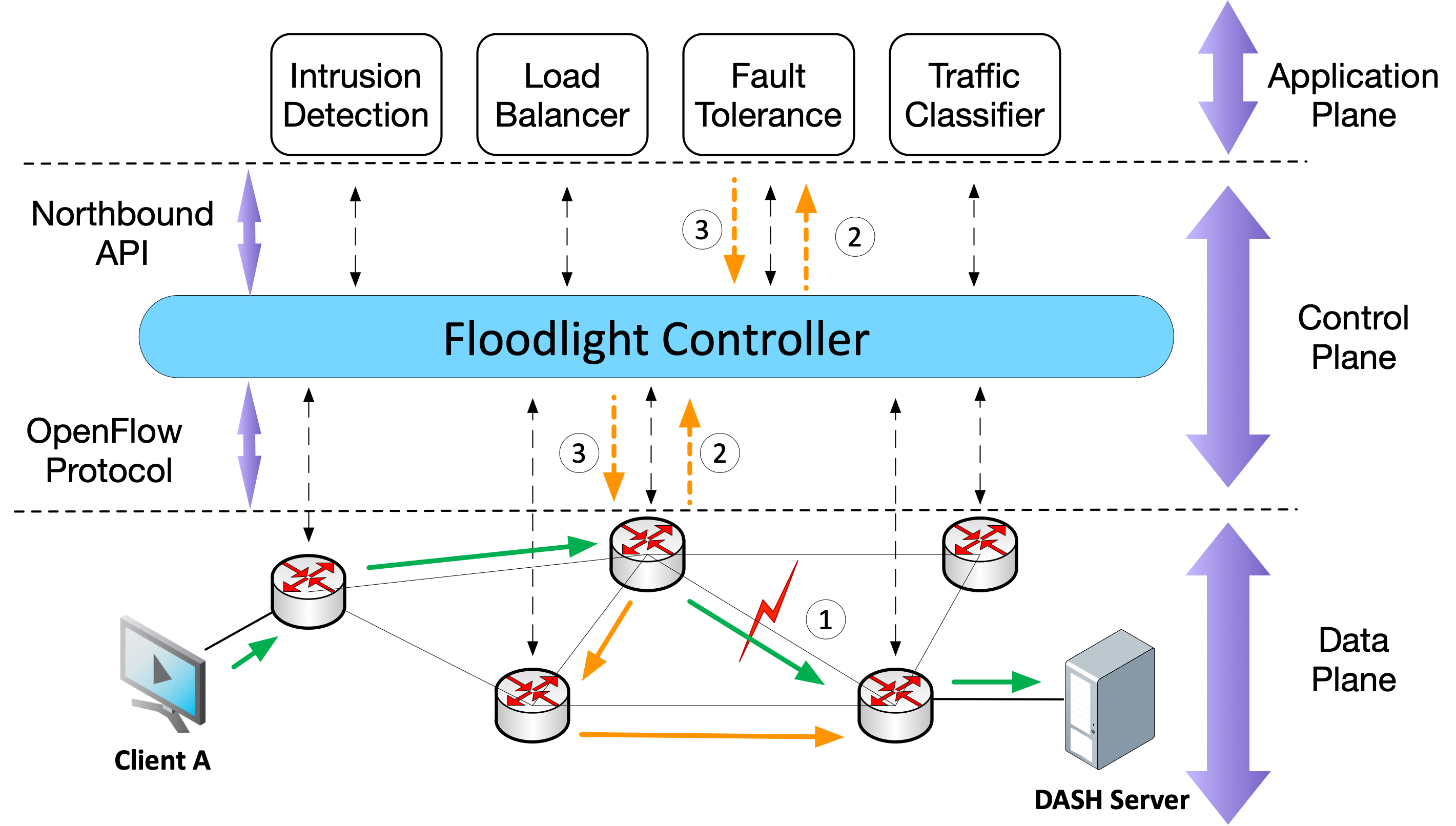}
 \DeclareGraphicsExtensions.
\caption{The design of the BFD-based Congestion Detection System.}
\label{systemDesign}
\end{figure}

We designed this module based on the data plane fault tolerance mechanism applying the restoration approach rather than protection. Since video streaming can resist network changes for seconds through its buffering mechanism, rerouting flows based on the recent network view would be beneficial. Similar to the fault tolerance problem, our BFD-based Congestion Detection Module includes a detection phase and the action phase. However, since there is no need to reroute all affected flows in the congestion problem, it is separated from the fault tolerance. Figure \ref{systemDesign} shows the main aspects of our design. 

\subsubsection{Congestion Detection}

The first step is the detection of the congestion using the BFD in the case of a heavy traffic load as shown in Figure \ref{systemDesign}. BFD is run on the switches to which the observed link is connected. These two switches are called as a pair regarding to BFD. Each switch conveys a control packet including the current state of the monitored link to its pair switch. When a switch receives the control message from its pair, it sends an echo message to it with the session status. 


The failure detection time, $T_d$, is based on the BFD message transmit interval, $T_i$, that can be manually configured by the network administrator considering the given services. For example, if real-time applications such as VoIP are widely used in the network, the interval must be very low considering the 50ms recovery time \citep{sharma2013openflow}. On the other hand, if there is multimedia traffic like video streaming, the interval may be in seconds due to the buffer mechanism of the applications. Thus, $T_d$ is computed based on the $T_i$ and the detection time multiplier $M$ as given in Equation \ref{eq2}. $M$ value is used to prevent false positives.

\subsubsection{Rerouting Flows}

After detecting the congestion using BFD, the problem on the link is reported to the controller via the OpenFlow protocol. Subsequently, the controller informs the modified fault tolerance application about the problem. In the application, flows passing through that link are first extracted from the flow pool in which all active flows are held. Afterwards, the predefined percentage of flows, which is 50\% in this study, are selected for rerouting and the new rules are created for them. As a result, the congested link is relieved.

\subsection{Restoration Module}
In the restoration, the essential idea is the involvement of the controller. Because of the dynamic nature of the network environment, the restoration approach is important to handle the failures in the network. We implemented the restoration module as demonstrated in Algorithm \ref{algres} by applying the steps shown in Figure \ref{restoration}. First, the failed link information is received by the controller as an event/input. Based on this information, the current network state is computed in order to calculate the new route based on the smallest hop count for the flows affected by the failure. Afterwards, the affected paths are identified via the active flow pool using the failed \textless switch, port\textgreater{} tuple and then a new path is calculated for each flow. Finally, old rules are replaced with the new ones determined by the restoration module.

\subsection{Static Protection Module}

\begin{figure}
\centering
\begin{subfigure}{0.45\textwidth}
\includegraphics[width=\linewidth]{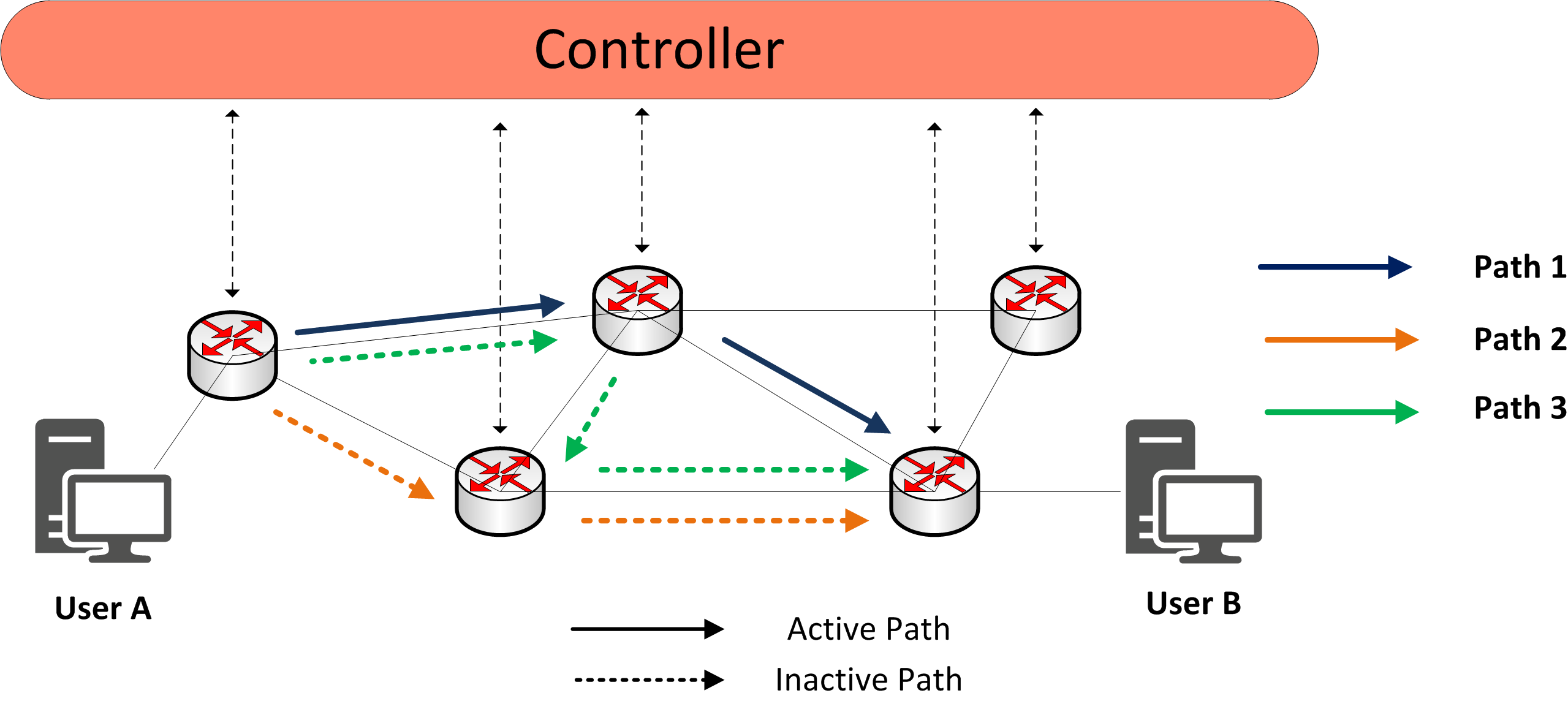}
\caption{Primary and backup paths between communicators.} \label{PrimaryBackupPaths}
\vspace*{0.5cm}
\end{subfigure}
\centering
\begin{subfigure}{0.45\textwidth}
\includegraphics[width=\linewidth]{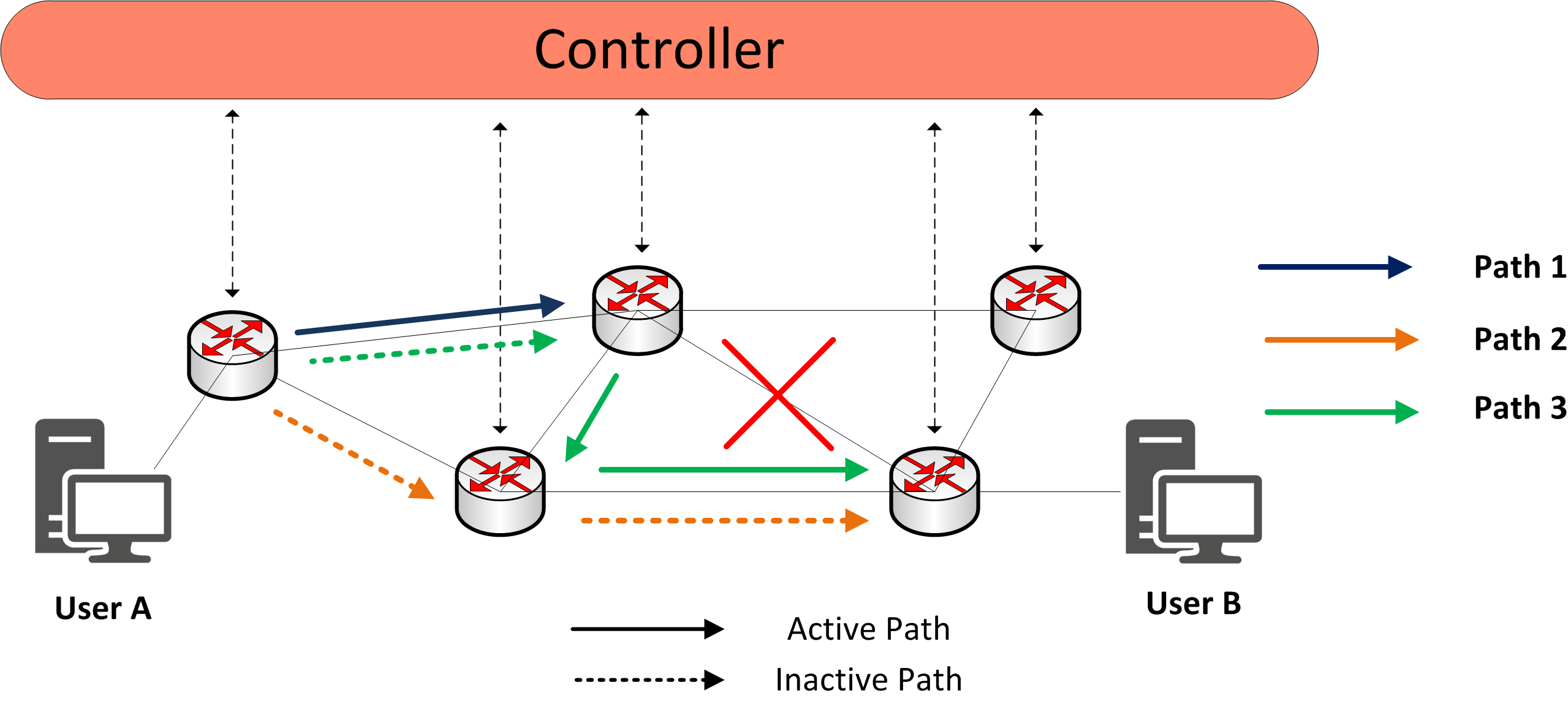}
\caption{Static protection using OpenFlow groups.} \label{StaticProtectionUsingOF}
\end{subfigure}
\caption{Static Protection in SDN.} \label{StaticProtectionSDN}
\end{figure}

We used OpenFlow Fast Failover groups for this module to apply backup paths in case a failure occurs in the primary path. Thus, the failure is recovered without the involvement of the controller and the burden on the control plane and control channels is mitigated as depicted in Figure \ref{StaticProtectionSDN}. First, all possible routes between the new source-destination pair are calculated and one of them is selected as the primary path based on the hop count. Afterwards, if a failure occurs at one of the resources on the primary path, Fast Failover groups handle it using the working buckets in which the backup actions are defined. As shown in Figure \ref{StaticProtectionUsingOF}, the primary path is actually used to forward packets until the switch to which the failed link is connected. However, since the watch/port group belonged to the primary path in that switch cannot work for the rest of the route because of the failure, the secondary path would become active by forwarding the packets of flows to the corresponding switch. Moreover, since the whole operation after the failure is carried out in the data plane, the burden on the controller would significantly reduce. 

Since the controller is not involved for the failure in the protection approach, this module considers only the installation of the flow rules including the primary and backup paths on the data plane in a proactive way. Hence, Algorithm \ref{algstatic} shows only the steps that how the routes are calculated initially and then the configuration with the installation of the Fast Failover groups. First, the current network state is computed for effective routing. Second, for a given source and destination pair, all possible paths between them are calculated and then assigned to the $allPaths$ variable. Afterwards, for each path in this set, flow rules are installed into the appropriate switches in the data plane by using $InstallProactiveRules$ function. In this function, since a path consists of \textless switch, port\textgreater{} tuples in our design, we investigate each switch in the path considering that whether it has the bucket that directs flows to the corresponding egress port or not. If it has buckets for given $groupID$, which indicates the group of paths between a \textless source, destination\textgreater{} tuple, and there is no egress port information in the bucket, we append it and then update the switch. On the other hand, if there are no buckets, we create one with the egress port information for the switch. Thus, all paths are installed into the data plane as the proactive procedure.

\begin{algorithm}[h]
\KwIn{\textless source, destination\textgreater{} tuple}
\KwOut{Primary and backup paths in the data plane using Fast Failover groups}
\SetKw{KwBy}{by}
\SetKwProg{Fn}{Function}{}{end}
\Fn{BuildingProtection()}{
ComputeCurrentNetworkState()\;
$allPaths$ = ComputeAllPaths($src$, $dst$)\;
$groupID$++\;
\For{$path \in allPaths$}{
	InstallProactiveRules($path$, $groupID$)
}

}

\Fn{InstallProactiveRules($path$, $groupID$)}{
$index$ = $path$.length()\;
\For{$switch \in path$}{
	$outPort$ = $path$.get($index$)\;
	$key$ = getKey(switch + groupID)\;
	$buckets$ = getBucketList($key$)\;
	$isBucketsExist$ = false\;
	\If{$buckets$ != null}{
		$isBucketsExist$ = true\;
	}
	\If{$isBucketsExist = false$}{
	$buckets$ = createBucketList()\;
	}
	\If{$buckets$.contain($outPort$) = false}{
			$buckets$.add($outPort$)\;
		}
	$switch$.update($buckets$)\;
	
}

}

\caption{Static Protection Module}
\label{algstatic}
\end{algorithm}

\section{Performance Evaluation of Network-based Metrics}

In this section, we evaluate the network-based metrics for QoS in fault tolerance. We conducted several experiments using iPerf \citep{iperf} and Wireshark \citep{wireshark} tools on several scenarios. To this end, we first evaluated the performance of our DPQoAP module comparing it with the traditional protection approach. Afterwards, we focused on the creation of the link failure deploying Linux bridge in Mininet to observe its impact on our fault-tolerant modules. Accordingly, we analyzed the effect of BFD intervals on the packet loss.

We repeated experiments 10 times for the variance control. We used Mininet for the SDN emulation, and Open vSwitch \citep{openvswitch} for virtual switches created in Mininet. 

In the experiments, based on the scenario, we trigger the failure in two ways:

\begin{enumerate}

\item \textit{Using Mininet Command:} We used the standard Mininet command, as \textit{link switchA switchB down}, to create the link failure in a given topology. However, in addition to the link, this command also destroys the ports of switches to which the link is connected. Since studies that investigate the fault tolerance problem in the data plane using Mininet execute this command for the failure, we also applied it in our experiments. 

\item \textit{Using Linux Bridge:} In the real world, failures on the link usually happen without affecting the ports of the switches like a broken cable. This is crucial since affected ports cause an immediate notification for the switch related to the failure. Thus, to emulate a real world like failure event in Mininet, we shut down one of the ports of the Linux bridge that is placed between the corresponding switches.

\end{enumerate}  

\begin{figure}
\includegraphics[scale=0.35]{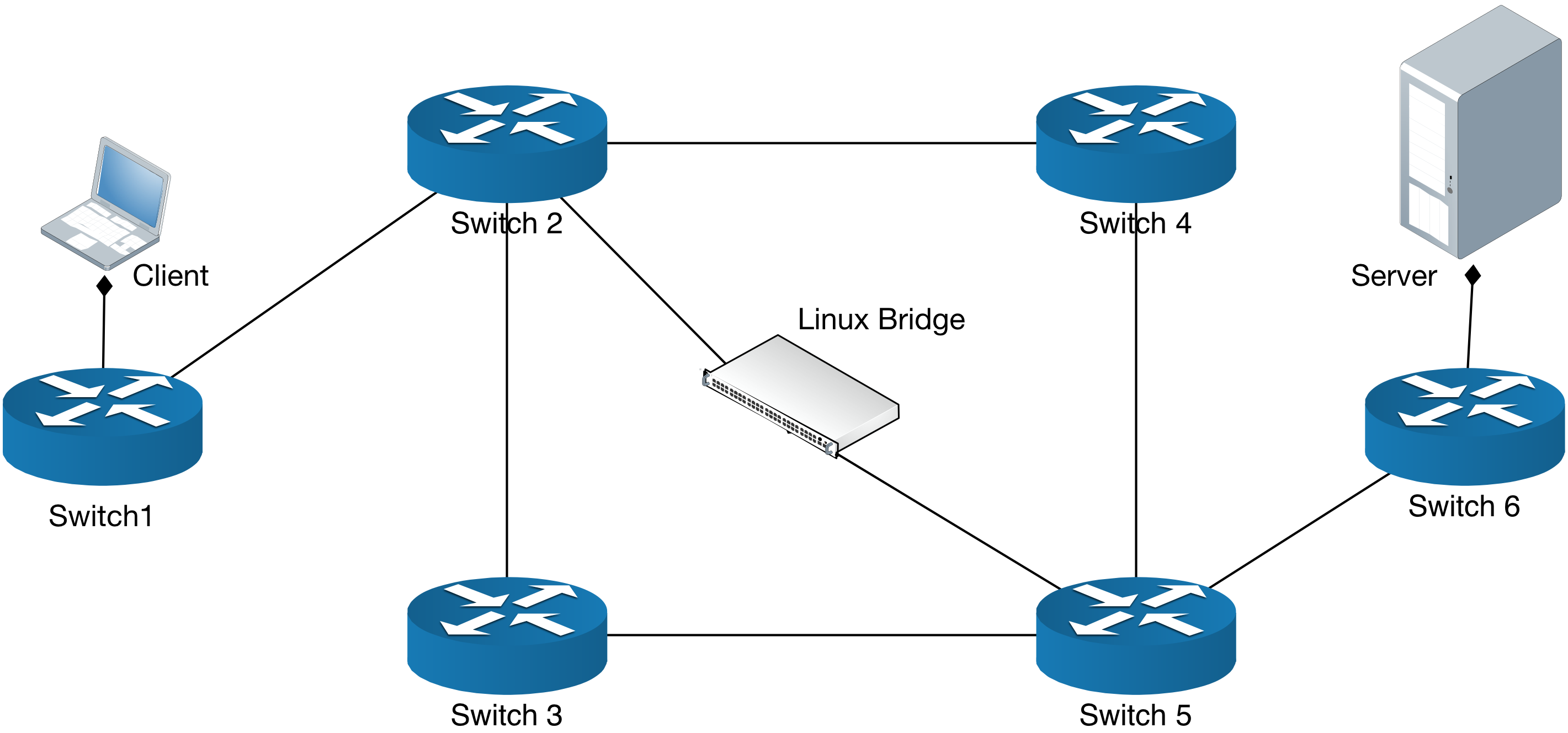}
\centering
 \DeclareGraphicsExtensions.
\caption{The topology using a Linux bridge.}
\label{topology-1}
\end{figure}

Considering the Linux bridge command, we used the topology depicted in Figure \ref{topology-1}. We also employed the same topology without deploying Linux bridge for the usage of the Mininet command.

\subsection{Evaluation of the DPQoAP}
\label{QoAP-1}

For a proper assessment of DPQoAP, we compare our module with the static protection module. In the test scenario, we create a heavy traffic on the secondary path before the link failure. Since affected flows are sent to the secondary path without considering the network conditions in the static protection, this comparison presents the benefits of the dynamic protection.

In our experiments each of which lasts 50 seconds, we used the topology shown in Figure \ref{topology-1} without deploying Linux bridge. In the topology, the primary path is S1-S2-S5-S6 while the secondary path is S1-S2-S3-S5-S6. We first created the traffic that causes 100\% load on the link between Switch 2 and 3 at the 10th second. Afterwards, at the 26th second, we applied the Mininet command to create the link failure on the link between Switch 2 and 5.

\begin{figure}[t]
\centering
\includegraphics[scale=0.20]{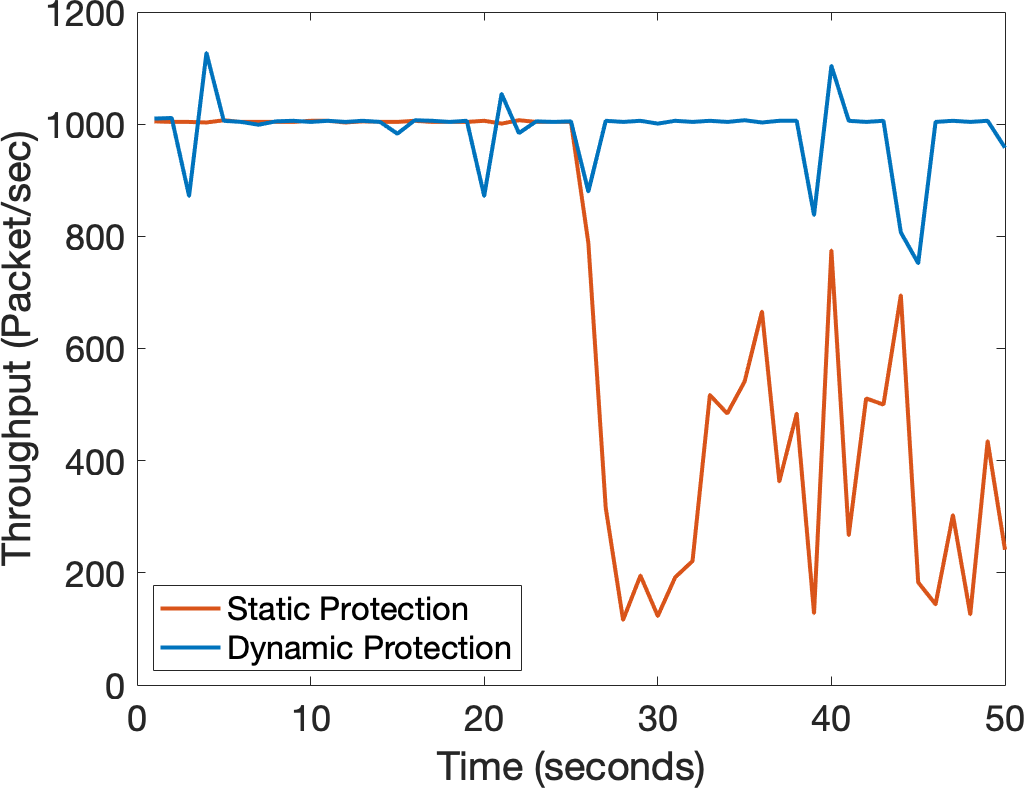}
 \DeclareGraphicsExtensions.
\caption{A throughput difference between the static and dynamic protection approaches considering QoAPs.}
\label{DynamicProtectionExperiment}
\end{figure}

The result shown in Figure \ref{DynamicProtectionExperiment} depicts that the throughput of the transmission significantly reduces when the QoAP is not considered. Moreover, using our DPQoAP module, the traffic is not affected by the conditions of the original secondary path after the link failure since all affected flows have been directed into a different route.

\begin{figure}
\centering
\includegraphics[scale=0.17]{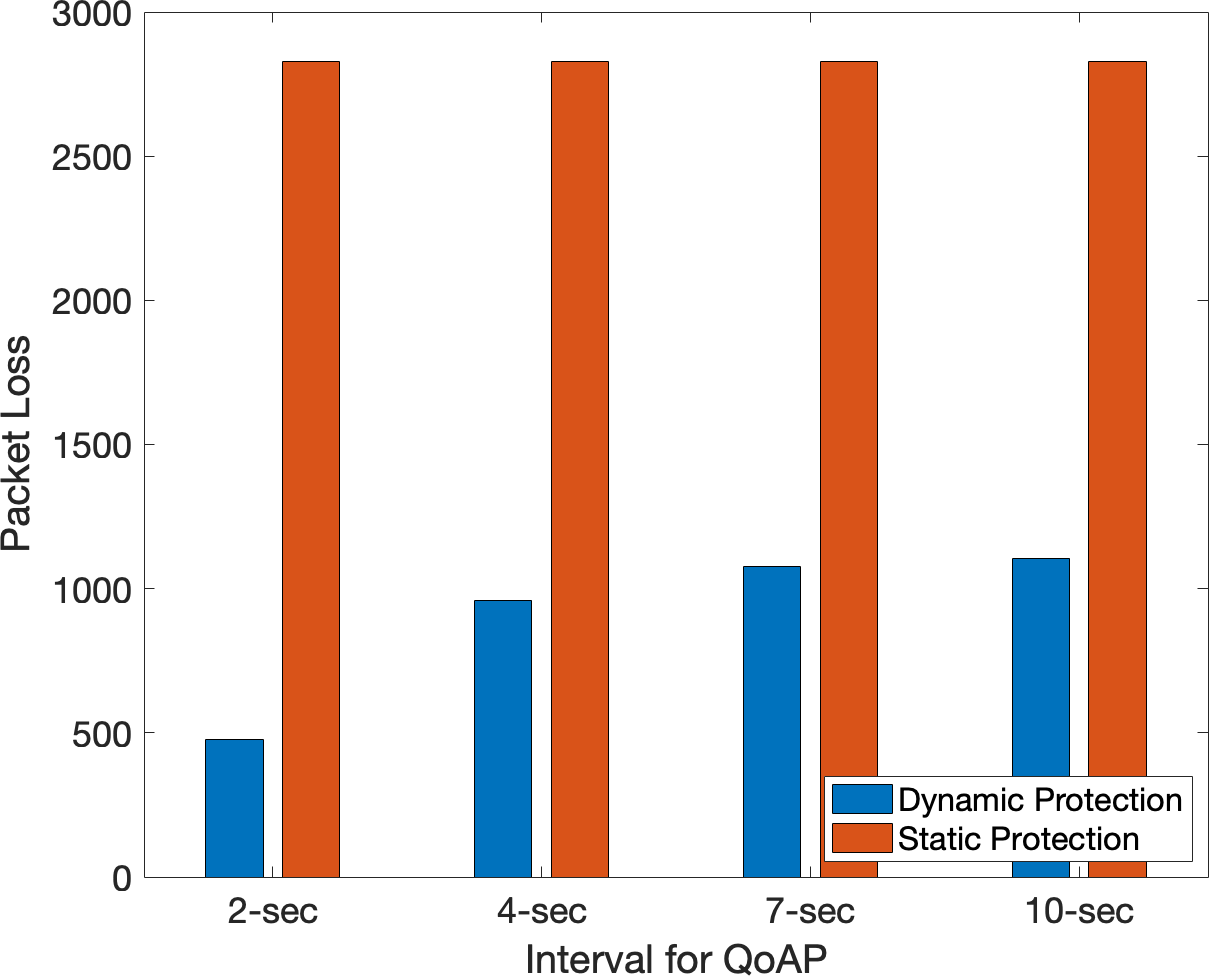}
 \DeclareGraphicsExtensions.
\caption{Comparison of the dynamic and static protection approaches based on the packet loss for each QoAP interval.}
\label{DynamicProtectionIntervals}
\end{figure}

On the other hand, evaluation of the interval parameter that is used to calculate the QoAPs for the given topology is crucial for the performance of the system. Therefore, we compared the performance of 2-sec, 4-sec, 7-sec, and 10-sec intervals considering the packet loss. Moreover, we also compared these results with the performance of the static protection. The results shown in Figure \ref{DynamicProtectionIntervals} presents that the performance of our dynamic protection approach outperforms the static protection. Since QoAPs are not considered in the static protection, packet losses are higher than the dynamic protection. Besides, the results also show that if the interval in DPQoAP decreases, the packet loss reduces since the application uses more recent information of the network. However, lower interval causes higher cost for the system since the statistics of the all living ports in the network must be collected within the given interval via LLDP packets. Thus, the interval must be optimized considering the controller traffic and the applications that run on the network.

\subsection{Evaluation of Linux Bridge and BFD Intervals}

To evaluate the effect of the Linux bridge command in order to observe the broken link scenario in Mininet, we first compared the transmission patterns of the link failures carried out by Mininet command and Linux bridge command, which consists of shutting down a port of Linux bridge, respectively. We applied the restoration and static protection approaches for the evaluation. We used the topology shown in Figure \ref{topology-1} for the Mininet-based failure and Linux bridge-based failure, respectively. In the experiments, we used a single flow to observe the Linux bridge effect independently. The duration of each experiment was 50 seconds in which the link failure was created at the 15th second.

\begin{figure}[!t]
\centering
\begin{subfigure}{0.30\textwidth}
\includegraphics[width=\linewidth]{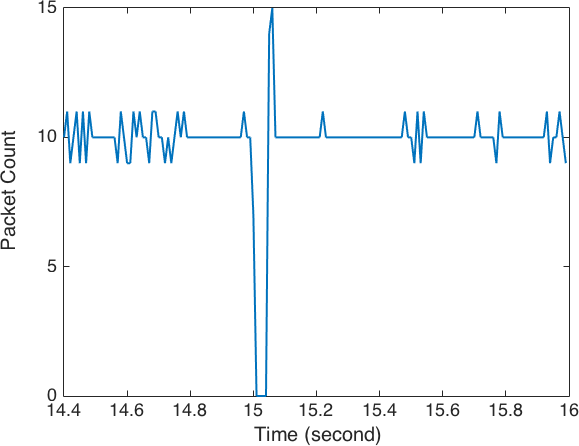}
\caption{Recovery time is 30-50ms for restoration module using Mininet command.} \label{MininetRecovery}
\end{subfigure}
\begin{subfigure}{0.30\textwidth}
\includegraphics[width=\linewidth]{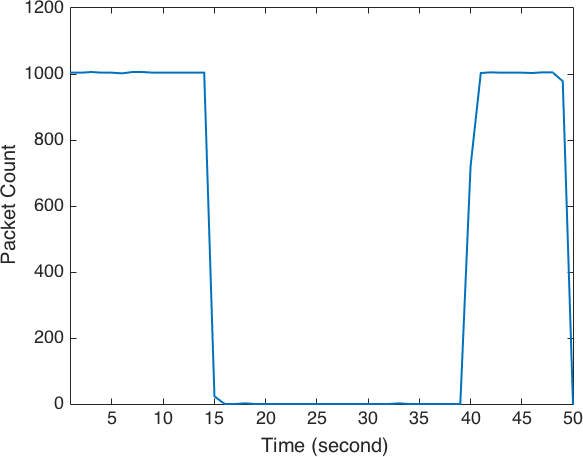}
\caption{Recovery time is in seconds for restoration module using Linux bridge for failure.} 
\label{LinuxBridgeRecovery}
\end{subfigure}
\caption{The demonstrative examples of failure recovery for Mininet command and Linux bridge based scenarios.} \label{DemonstrativeExamples-1}
\end{figure}

Considering the link failure carried out by the Mininet command, Figure \ref{MininetRecovery} shows the pattern of the transmission for the restoration approach based on the packet count. This pattern also shows the recovery time of the restoration approach whose mean is 40ms. On the other hand, if we shut down one of the ports of the Linux bridge to create the link failure, the failure notification event cannot be created immediately due to ports are active and therefore the controller cannot be notified. Thus, in this case, the controller realizes the link failure in the data plane via LLDP packets. Since LLDP packets are sent by the controller periodically for the topology discovery update in seconds to prevent the burden on the controller, the failure detection time is much higher than the other methods. Moreover, the failure detection time using LLDP updates in the Floodlight controller is twice of the LLDP update time, which is 12 seconds in our module. Thus, the pattern of the throughput in this case is as shown in Figure \ref{LinuxBridgeRecovery}. However, if we use the BFD protocol between Switches 2 and 5 using $T_i$ as 5ms, the mean of the recovery time would become 27ms that is independent of the failure creation approaches including Linux bridge and Mininet commands. 

\begin{figure}[t]
\centering
\includegraphics[scale=0.20]{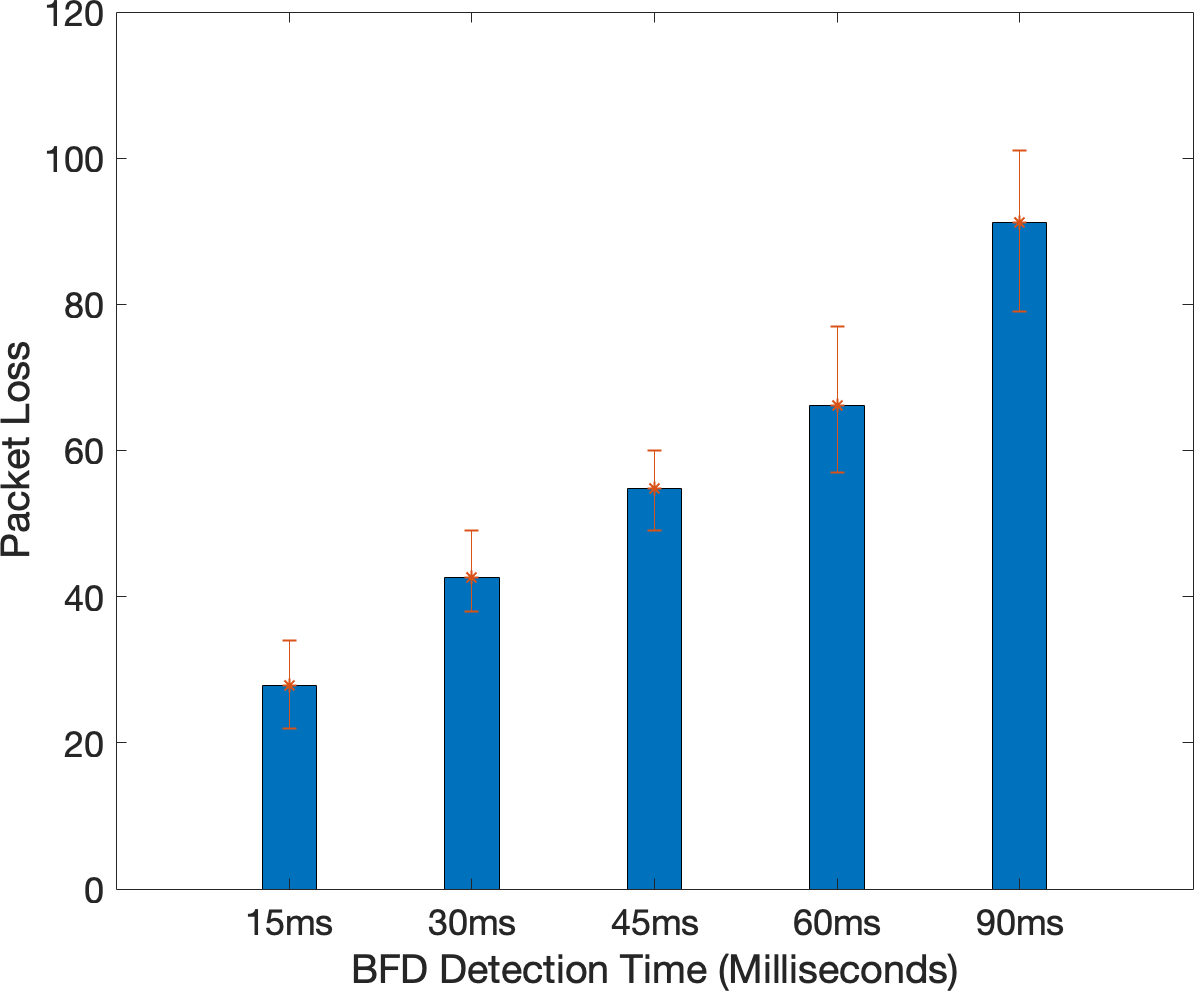}
 \DeclareGraphicsExtensions.
\caption{Packet Loss based on the BFD message interval.}
\label{changingBFD}
\end{figure}

We applied the same approaches above for the evaluation of the static protection module. Since this module solves the problem in the data plane using OpenFlow Fast Failover groups, the recovery time for the link failure created by the Mininet command is smaller than the restoration module with the mean value of 28ms. However, on the other hand, if we shut down the port of the Linux bridge for the failure, the transmission of the packets halt. Since the relevant ports of Switch 2 and 5 are not affected by the failure in this case, OpenFlow FF groups continue to operate due to it checks the liveliness of the ports even though there is a fault on the link between them. Thus, packets are sent to the ports that are assumed as working and then the traffic stop. However, if we run the BFD protocol on that link, the transmission continues flawlessly since Open vSwitch works based on the value of $T_i$.

Since we have observed through our experiments that BFD is crucial for a fault tolerant system, we also evaluated the effect of different $T_i$ values on the recovery. We considered 15ms, 30ms, 45ms, 60ms, and 90ms failure detection times regarding Equation \ref{eq2} and measured the packet loss. The result shown in Figure \ref{changingBFD} presents that when $T_i$ increases, the packet loss also increases since the duration of the failure detection is prolonged.

\section{Performance Evaluation of Application-based Metrics}

Evaluating the change of the application-based metrics is crucial as well as assessing the network-based parameters since end-users are affected in this case. To this end, we focused on the change of QoE parameters in two ways: (1) when the failure happens and (2) when there is a congestion in the network.

\subsection{The Effect of Link Failure on QoE}

Our scenarios for this evaluation are similar to the network-based assessment considering the performance of the restoration, static protection and DPQoAP modules in the case of a link failure. In the experiments, we used the topology shown in Figure \ref{topology-1} including one client and one server. The client is connected to Switch 1 and the server is connected to Switch 6. We used a short movie namely Big Buck Bunny with 1080p resolution for the video streaming. The length of the movie is 600 seconds and the link failure was carried out at the 300th second for each experiment. Moreover, we used 1 second and 10 seconds segment sizes in order to assess the effect of the segment size. We investigated the change of the video quality value and buffer level as the application metrics considering QoE. To measure the video quality value, we used Equation \ref{eq5} while the buffer level information was provided by DASH API. 

\begin{figure*}[!ht]

\begin{subfigure}{0.23\textwidth}
\includegraphics[width=\linewidth]{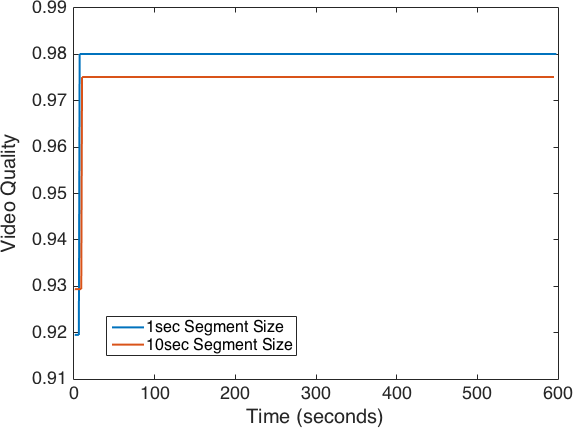}
\caption{Video quality value when we use Mininet command for the failure.} 
\label{VideoQualityForRestoration-1a}
\end{subfigure}
\hspace*{\fill} 
\begin{subfigure}{0.23\textwidth}
\includegraphics[width=\linewidth]{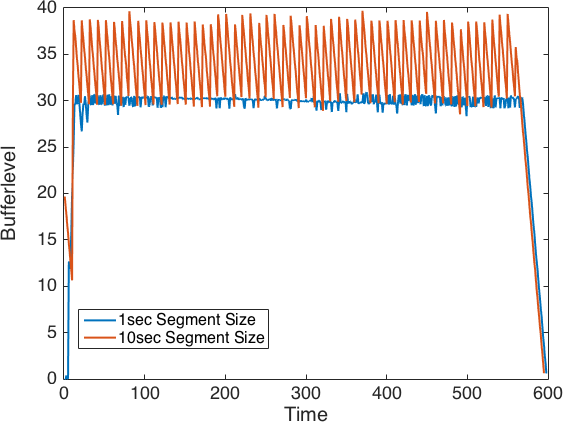}
\caption{Buffer level of the video when we used Mininet command for the failure.} \label{VideoQualityForRestoration-1b}
\end{subfigure}
\label{VideoQualityForRestoration-1}
\hspace*{\fill} 
\begin{subfigure}{0.23\textwidth}
\includegraphics[width=\linewidth]{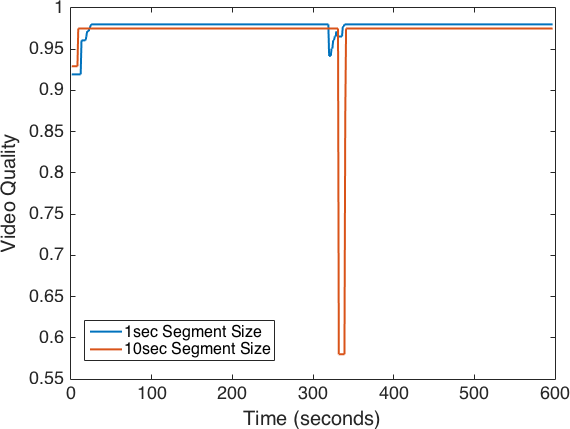}
\caption{Video quality value is affected when we use Linux bridge for the failure.} \label{VideoQualityForRestoration-2a}
\end{subfigure}
\hspace*{\fill} 
\begin{subfigure}{0.23\textwidth}
\includegraphics[width=\linewidth]{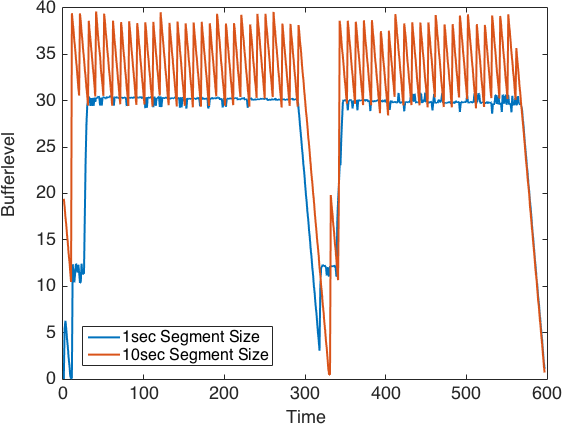}
\caption{Buffer level is affected when we use Linux bridge for the failure.} 
\label{VideoQualityForRestoration-2b}
\end{subfigure}
\caption{The effect of failure recovery on the video quality and buffer level of video streaming using restoration module with Mininet-based and Linux bridge-based failure.}
\label{VideoQualityForRestoration-2}
\end{figure*}

\subsubsection{Restoration}

When the restoration module is used, the video quality and buffer level is not affected by the failure created by the Mininet command for both segment sizes as shown in Figure \ref{VideoQualityForRestoration-1a} and \ref{VideoQualityForRestoration-1b}. Since video streaming is based on HTTP and the buffer can hold the fragments of the video for seconds, a failure, which is recovered within milliseconds, does not affect the quality.

On the other hand, when we use the Linux bridge to create the link failure, the buffer level and therefore the video quality is affected for seconds since the recovery time is based on the LLDP update interval period in this case. As a result, QoE of the user reduces for seconds as shown in Figure  \ref{VideoQualityForRestoration-2a} and \ref{VideoQualityForRestoration-2b}. Another important result is that the failure impacts the video consisting of 10 seconds segment size more than the 1 second segment size. The main reason is that filling a 10 seconds long segment after the failure requires longer time than that of the 1 second segment size. Thus, the video cannot be continued to play properly because of the DASH concept which requires a full segment for streaming.

\subsubsection{Static Protection}

Using the static protection module, the buffer level and video quality are not affected by the failure created by the Mininet command as well as the restoration module. However, if we use the Linux bridge to create the failure, the transmission of the packets halts as in the evaluation of QoS using iPerf. Thus, the video continues a few seconds after the failure until the buffer in the client is depleted. On the other hand, when we activate the BFD protocol on the faulty link, the buffer level and the video quality is not affected by the failure as shown in Figure \ref{VideoQualityForRestoration-1a} and \ref{VideoQualityForRestoration-1b}.

\subsubsection{DPQoAP}

\begin{figure}
\begin{subfigure}{0.23\textwidth}
\includegraphics[width=\linewidth]{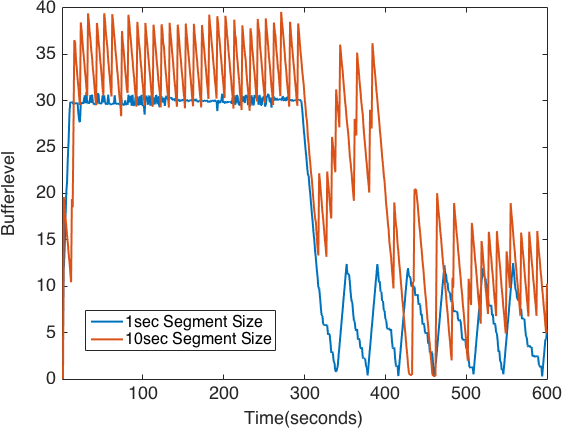}
\caption{Buffer level without considering QoAP.}
\label{QoAPforVideo-1a}
\end{subfigure}
\hspace*{\fill} 
\begin{subfigure}{0.23\textwidth}
\includegraphics[width=\linewidth]{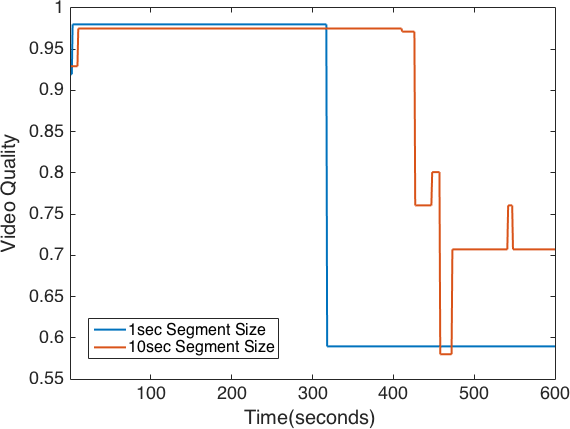}
\caption{Video quality value without considering QoAP.} 
\label{QoAPforVideo-1b}
\end{subfigure}
\caption{QoE is significantly affected if the quality of the alternative paths is not considered for fault tolerance.}
\label{QoAPforVideo}
\end{figure}

To evaluate the performance of the dynamic protection for the video application, we applied a similar scenario carried out in Section \ref{QoAP-1}. We created a heavy traffic on the secondary path at the 100th second before the failure. Afterwards, the link failure was created on the primary path, between Switch 2 and 5, using the Mininet command at the 300th second. We did not use the Linux bridge command for the failure in this case since our main consideration was to evaluate the quality of the communication after the failure rather than evaluating the recovery time. The result shown in Figure \ref{QoAPforVideo-1a} and \ref{QoAPforVideo-1b} points out that the QoE significantly reduces after the failure if we use the static protection in this scenario. On the other hand, if we use our DPQoAP module, QoE is not affected by the failure. Thus, it is clearly seen that considering the quality of alternative paths is also crucial for applications in a fault tolerant system as well as the recovery time.

\subsection{The Effect of Congestion on QoE}

We investigated three factors in our experiments for the congestion case: the impact of the BFD interval, the traffic load, and the video segment size on the QoE parameters. For each experiment, we used Mininet for SDN emulation deploying Open vSwitch for switches since it supports both BFD and OpenFlow protocols. On the other hand, we used DASH.js for the video clients. To generate additional traffic and thus cause congestion, we run the iPerf tool on Mininet.

\begin{figure}[!b]
\centering
\includegraphics[scale=0.065]{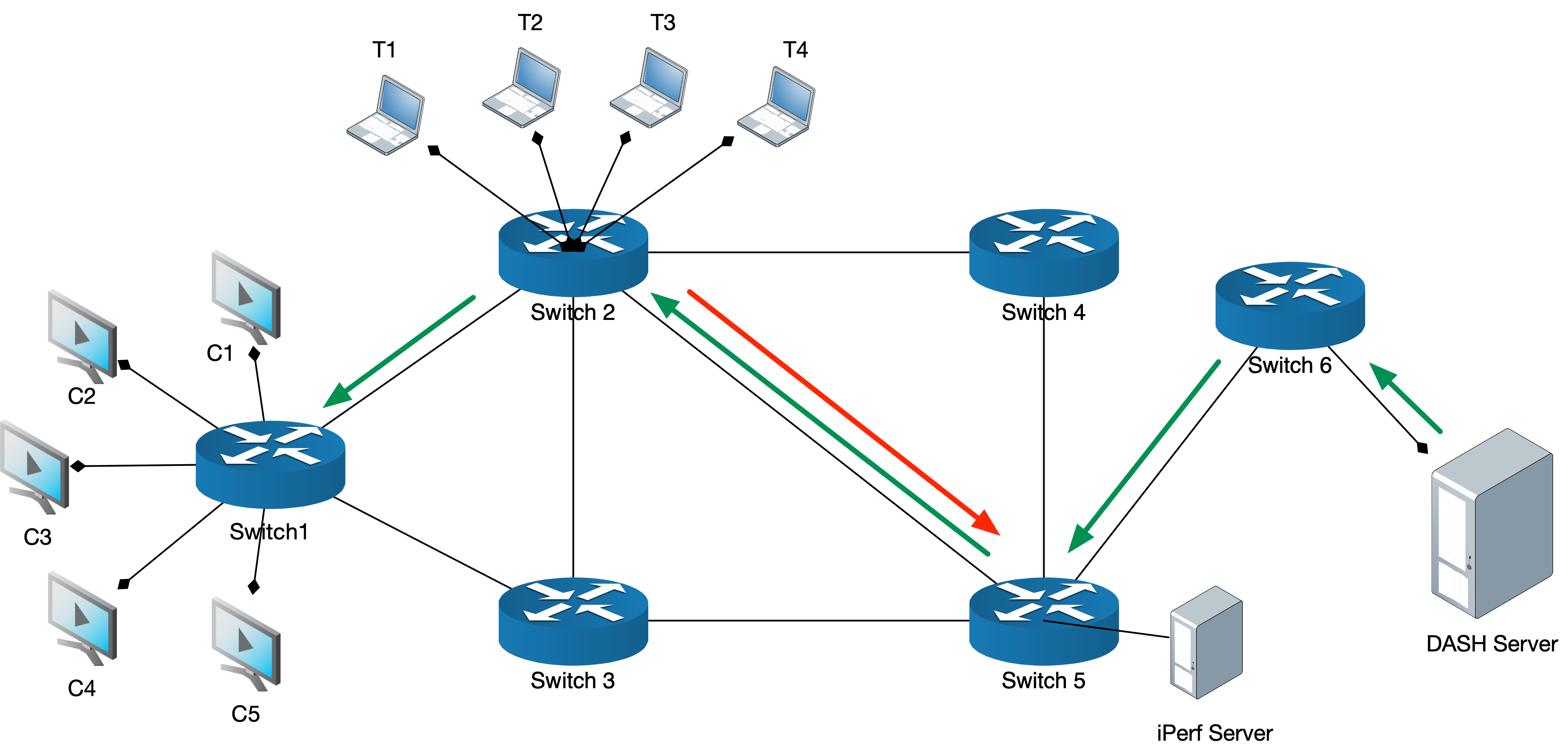}
 \DeclareGraphicsExtensions.
\caption{The topology and video traffic route}
\label{topology}
\end{figure}

\begin{figure*}[h]
\hspace*{\fill} 
\begin{subfigure}{0.23\textwidth}
\includegraphics[width=\linewidth]{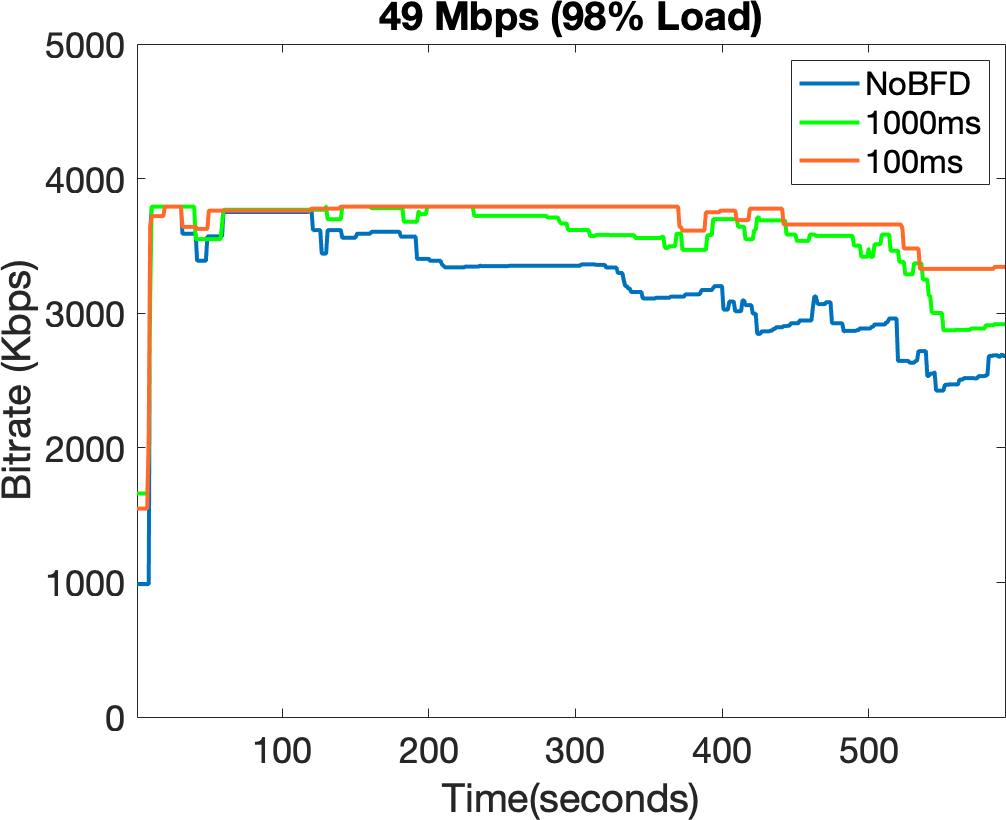}
\caption{Average bitrate for 10-sec segment size} \label{fig:1a}
\end{subfigure}
\hspace*{\fill} 
\begin{subfigure}{0.23\textwidth}
\includegraphics[width=\linewidth]{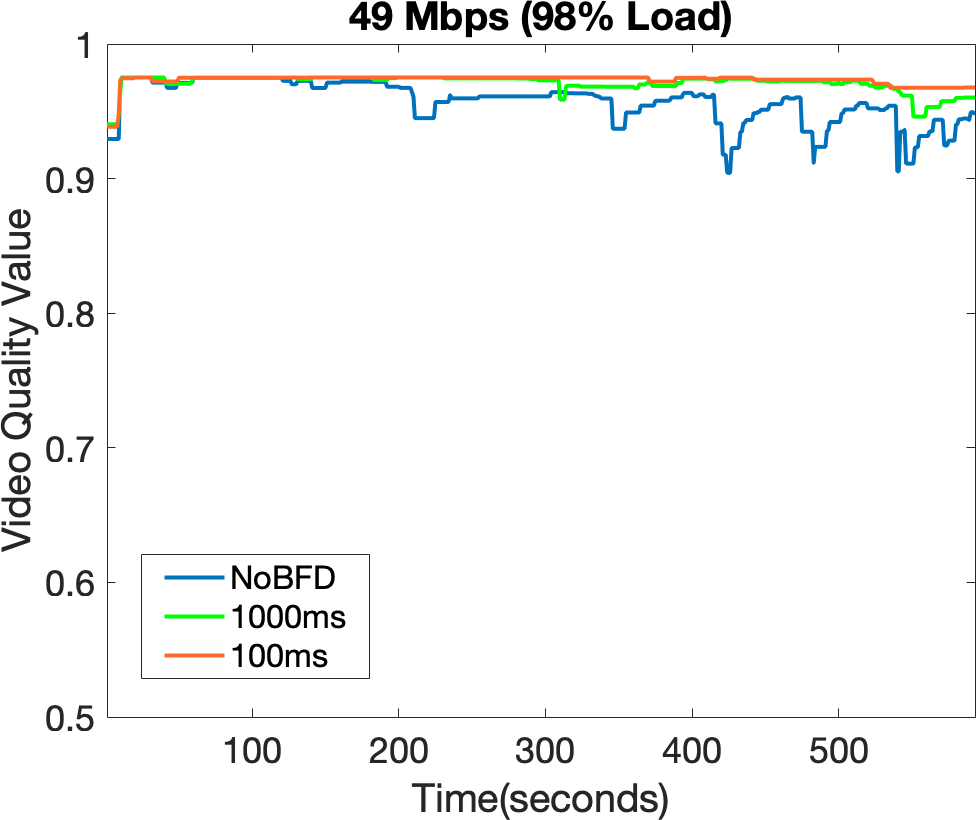}
\caption{Average video quality for 10-sec segment size} \label{fig:1b}
\end{subfigure}
\hspace*{\fill} 
\begin{subfigure}{0.23\textwidth}
\includegraphics[width=\linewidth]{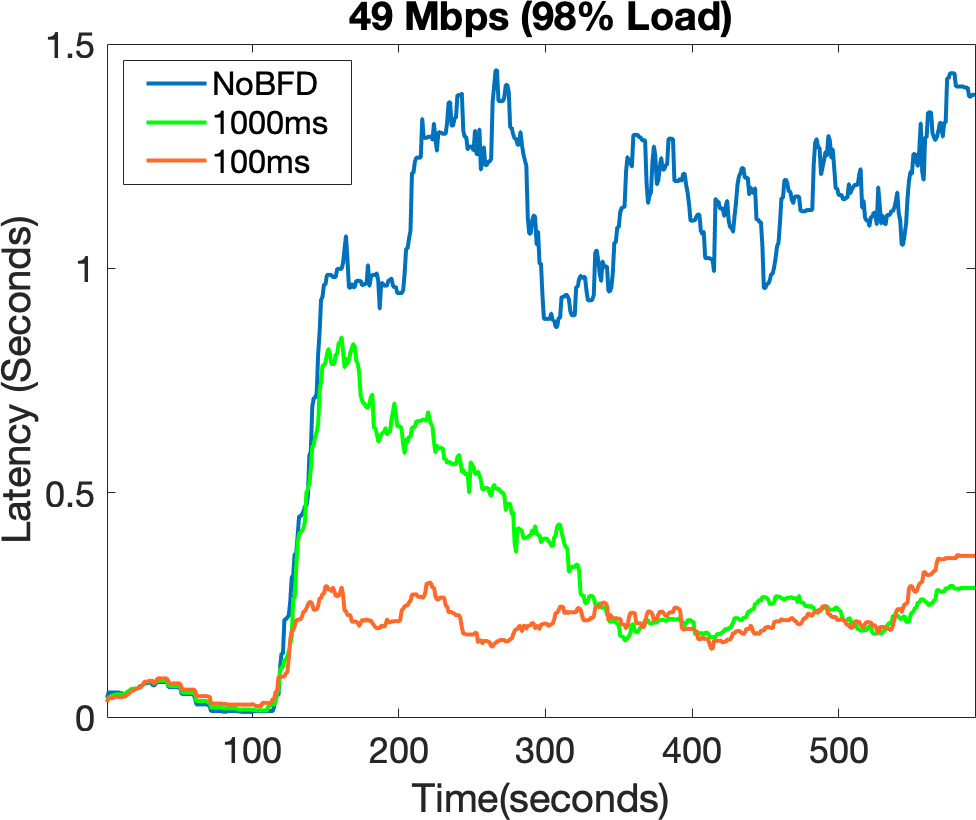}
\caption{Average latency for 10-sec segment size} \label{fig:1c}
\end{subfigure}
\hspace*{\fill} 
\begin{subfigure}{0.23\textwidth}
\includegraphics[width=\linewidth]{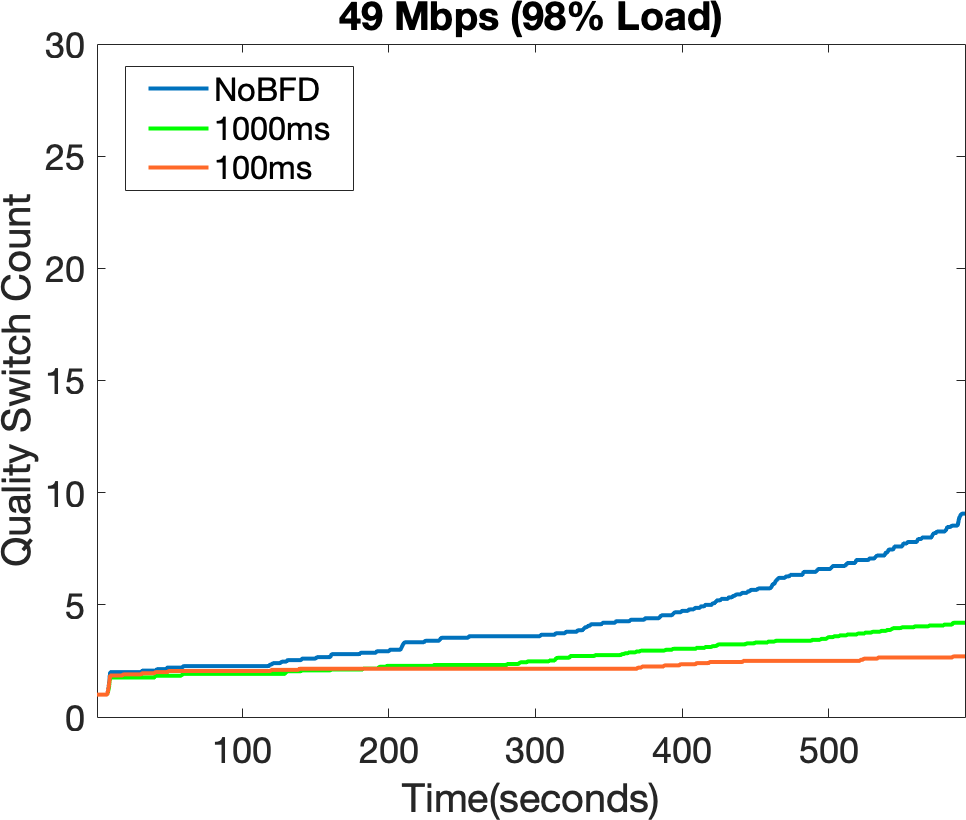}
\caption{Average number of quality switches for 10-sec segment size} \label{fig:1d}
\end{subfigure}

\caption{The change of QoE parameters for the congested link with 98\% load. The QoE parameters are affected by the congestion after 150th second. If our scheme is used with the BFD mechanism, the QoE parameters are improved for each case. }
\label{49MbpsTimeDomain-10sec}
\end{figure*}

\begin{figure*}[h]
\hspace*{\fill} 
\begin{subfigure}{0.23\textwidth}
\includegraphics[width=\linewidth]{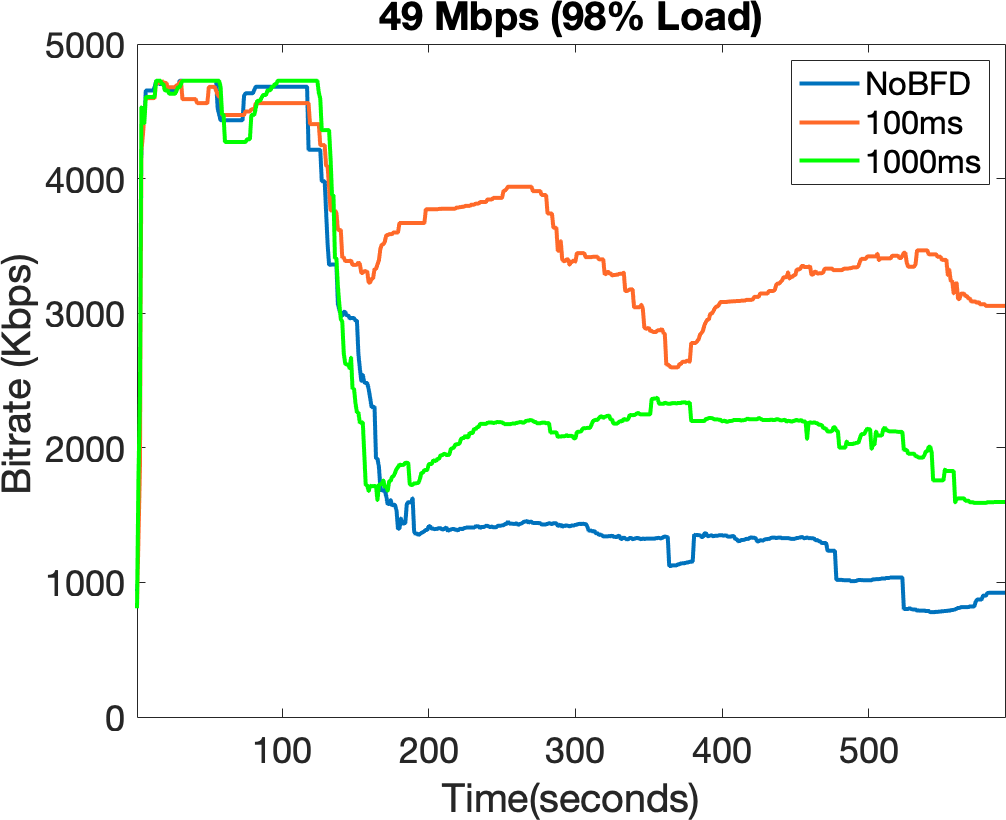}
\caption{Average bitrate for 1-sec segment size} \label{fig:2a}
\end{subfigure}
\hspace*{\fill} 
\begin{subfigure}{0.23\textwidth}
\includegraphics[width=\linewidth]{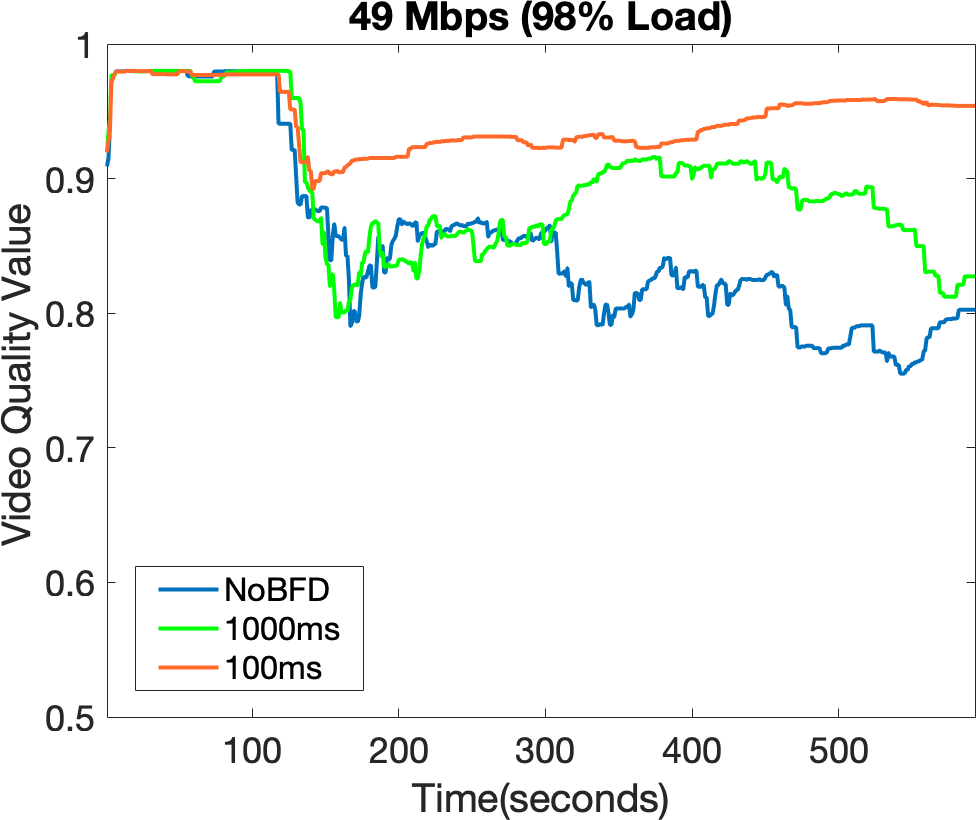}
\caption{Average video quality for 1-sec segment size} \label{fig:2b}
\end{subfigure}
\hspace*{\fill} 
\begin{subfigure}{0.23\textwidth}
\includegraphics[width=\linewidth]{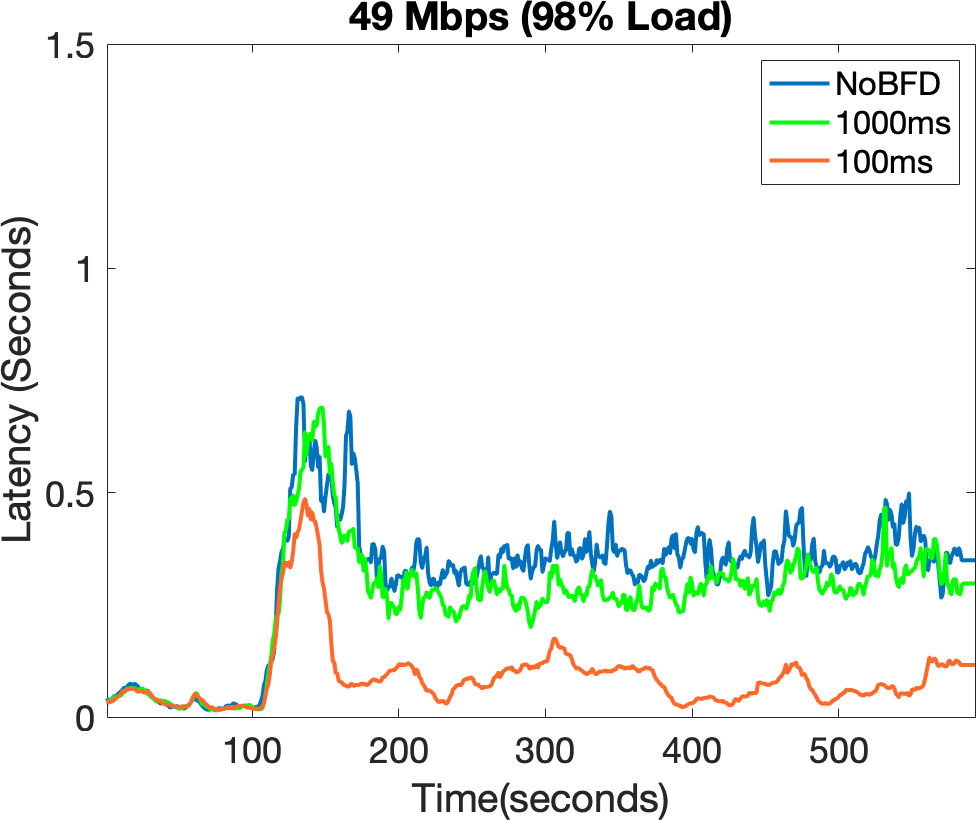}
\caption{Average latency for 1-sec segment size} \label{fig:2c}
\end{subfigure}
\hspace*{\fill} 
\begin{subfigure}{0.23\textwidth}
\includegraphics[width=\linewidth]{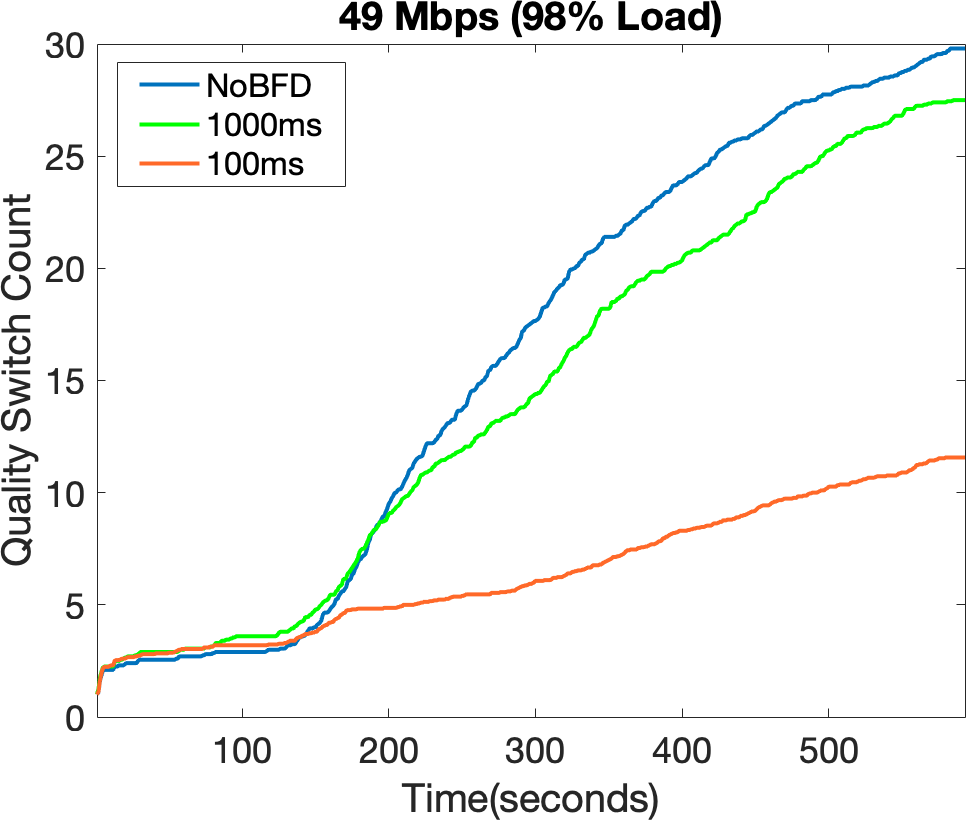}
\caption{Average number of quality switches for 1-sec segment size} \label{fig:2d}
\end{subfigure}

\caption{The change of QoE parameters for congested link with 98\% load. The QoE parameters are affected by the congestion after 150th second. If our scheme is used with BFD mechanism, the QoE parameters are improved for each case. }
\label{49MbpsTimeDomain-1sec}
\end{figure*}

For each experiment, the topology shown in Figure \ref{topology} was used. We limited the capacity of all links as 50Mbps by deploying five DASH clients connecting to Switch 1 and one DASH server attaching to Switch 6. We ensured that all DASH clients obtain the video segments via the same route as indicated in  Figure \ref{topology} as green arrows. On the other hand, four iPerf clients were connected to Switch 2 to cause congestion by sending their packets to the iPerf server which is located at Switch 5. Moreover, Big Buck Bunny short movie, the duration of which is 10 minutes, is used for streaming the 1080p resolution video. After the streaming was started for each DASH client, T1, T2, T3, and T4 iPerf clients began to send their packets at 50th, 80th, and 110th second respectively to cause congestion. These iPerf clients generated the same amount of UDP traffic and induced congestion on the link between Switch 2 and Switch 5.

We used 1 second and 10 seconds segments to evaluate the effect of the segment size. To observe the impact of the traffic load, we generated 40 Mbps (80\% Load), 45 Mbps (90\% Load), and 49 Mbps (98\% Load) traffic using only the iPerf clients. Finally, to evaluate the influence of the BFD intervals in our system, we used $T_i$ as 100ms and 1000ms respectively by taking $M$ value as two. Thus, the $T_d$ values were 300ms and 3000ms respectively. Moreover, we compared these results with the non-BFD case. 

Consequently, we conducted our experiments for 18 different cases considering those three parameters. For each case, we repeated experiments 6 times. Since each test lasts 10-11 minutes, the duration of our experiments was 18 hours. To evaluate the results, we analyzed four QoE parameters including the bitrate, quality value, latency and number of quality switches from each client and calculated the average values for each case.

\subsection{Effect of Segment Size}

Our experiments showed that streaming with the big segment size is more stable than the small segment size considering the congestion conditions on the link. Experimental results for 49 Mbps (98\% Load) traffic on the congested link given in Figure \ref{49MbpsTimeDomain-10sec}  and Figure \ref{49MbpsTimeDomain-1sec} represent that fluctuations and change of the QoE parameters including the average bitrate, video quality, latency, and number of quality switches between representations are higher in 1 second segment size compared with 10 seconds segment size. This pattern is the same for 45 Mbps and 40 Mbps traffic loads. Since a small segment size needs more HTTP requests to transmit video segments, it is affected by the network conditions more than the big segment sizes.

\subsection{The Effect of Traffic Load}

\begin{figure}
\begin{subfigure}{0.23\textwidth}
\includegraphics[width=\linewidth]{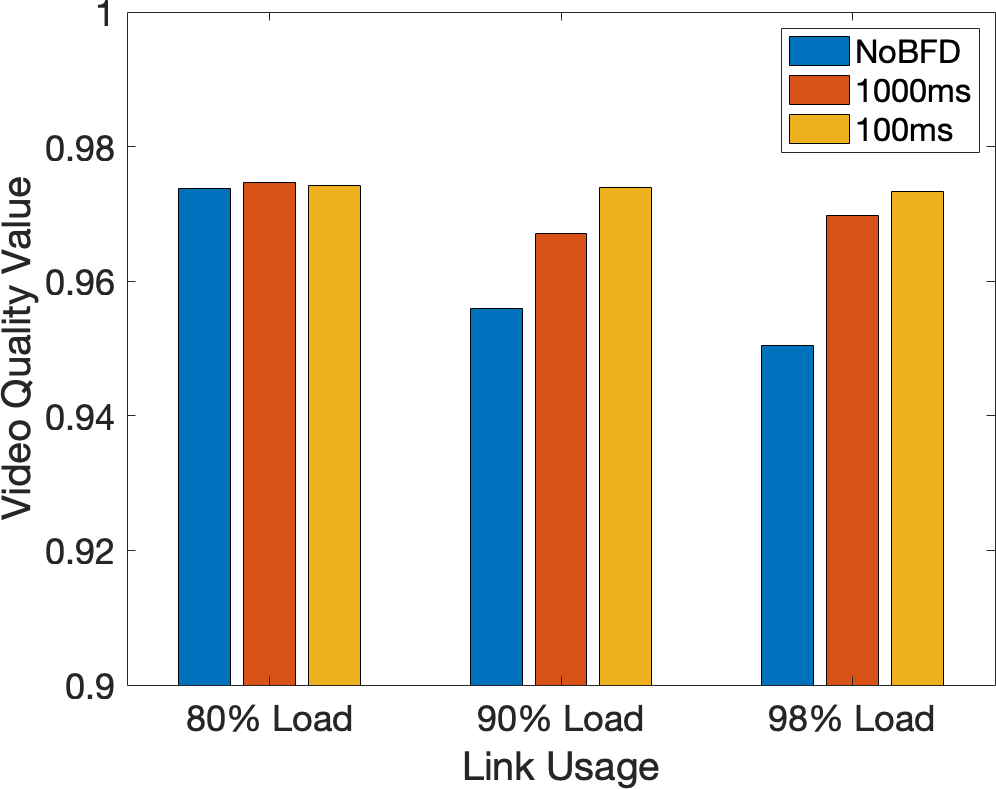}
\caption{The impact of the traffic load on the average video quality using 10-sec segment size} 
\label{ssim10sec}
\end{subfigure}
\hspace*{\fill} 
\begin{subfigure}{0.23\textwidth}
\includegraphics[width=\linewidth]{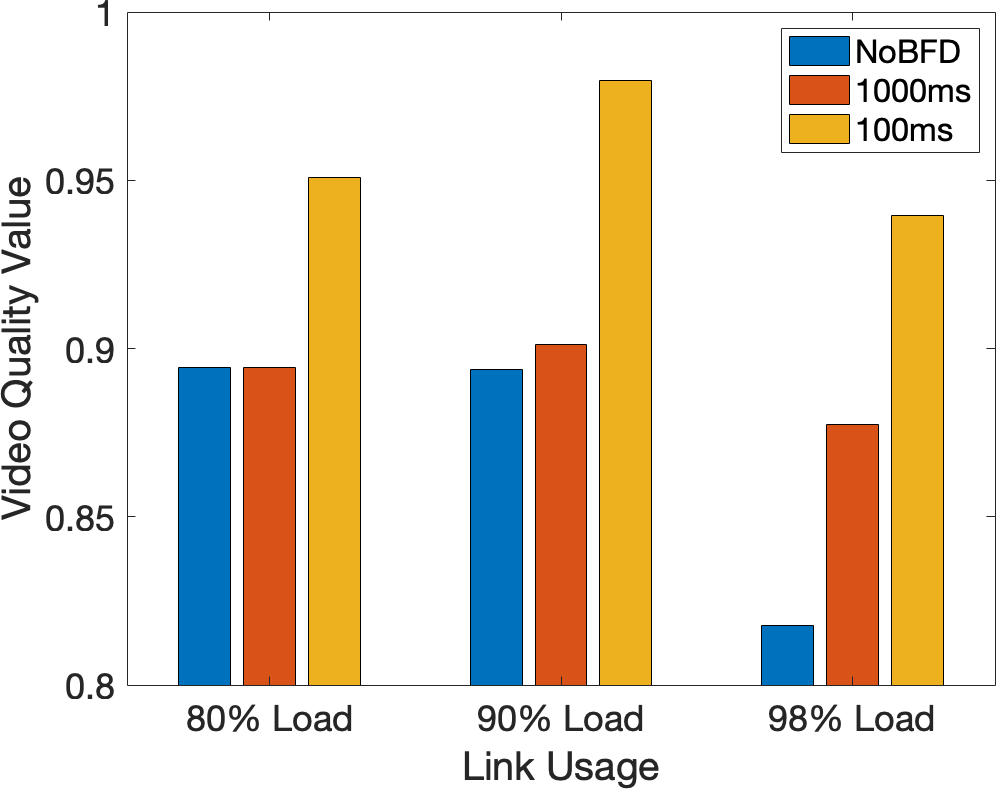}
\caption{The impact of the traffic load on the average video quality using 1-sec segment size} \label{ssim1sec}
\end{subfigure}
\caption{The impact of the traffic load on the average video quality with respect to different segment sizes}
\label{LoadToSSIM}
\end{figure}

To evaluate the impact of the traffic load, we measured the average video quality of clients based on SSIM and the number of quality switches including all cases in our experiments. Our results demonstrated that when the traffic load increases, the video quality decreases considering both segment sizes with the non-BFD case as shown in Figure \ref{ssim10sec} and \ref{ssim1sec}. For each traffic load, the video quality with 10-sec segment size is better than the 1 second segment size for the non-BFD case due to its buffer capacity. On the other hand, if we use BFD for the congestion detection, the video quality is improved.

Considering the non-BFD case for the 80\% and 90\% traffic loads, the average video quality with 1 second segment size is not so affected while the average video quality with 10-sec segment size is decreased. However, for the 98\% traffic load, the video quality is poor when we used 1 second segment size while it is acceptable for 10-sec segment size. 

\begin{figure}
\begin{subfigure}{0.23\textwidth}
\includegraphics[width=\linewidth]{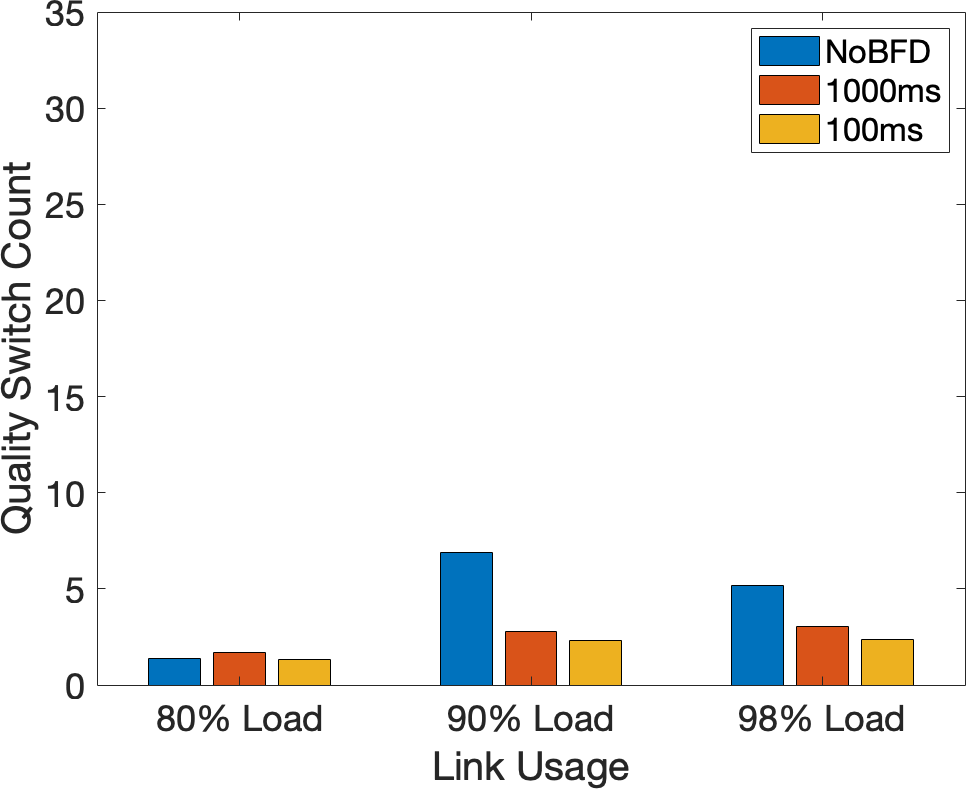}
\caption{The impact of the traffic load on the average quality switch count using 10-sec segment size} 
\label{switch10sec}
\end{subfigure}
\hspace*{\fill} 
\begin{subfigure}{0.23\textwidth}
\includegraphics[width=\linewidth]{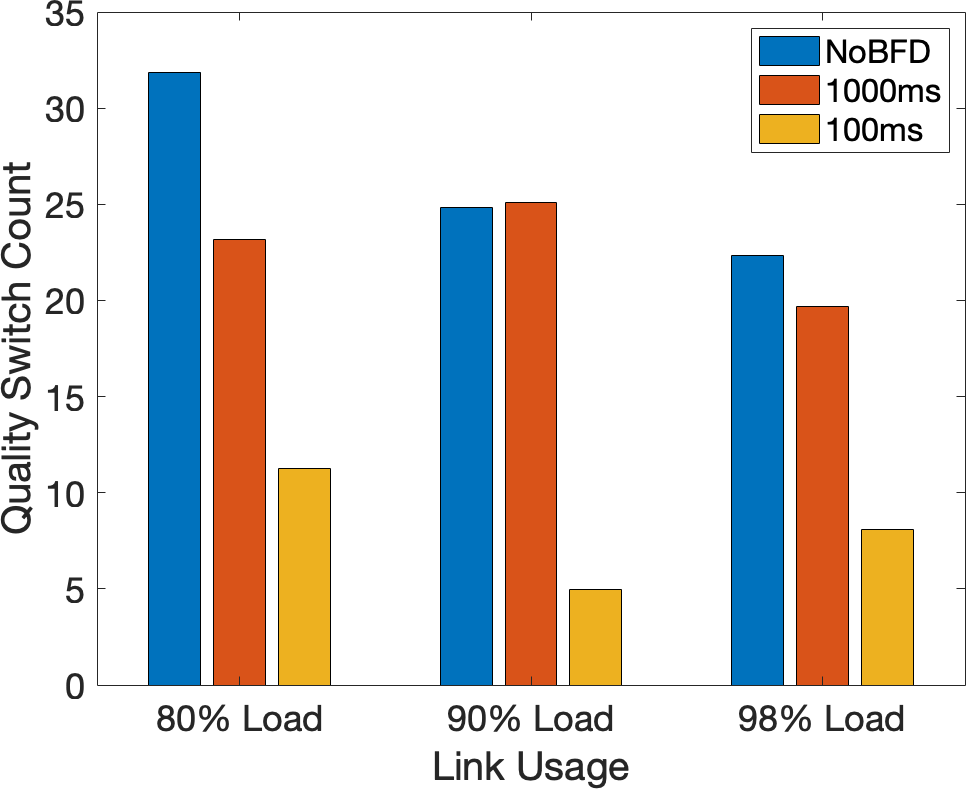}
\caption{The impact of the traffic load on the average quality switch count using 1-sec segment size} 
\label{switch1sec}
\end{subfigure}
\caption{The impact of the traffic load on the quality switch count with respect to different segment sizes.}
\label{LoadToSwitch}
\end{figure}

On the other hand, the effect of the traffic load on the number of quality switches is shown in Figure \ref{switch10sec} and \ref{switch1sec}. The results show that the quality switch count between representations for the 10-sec segment size is the lowest for 80\% load and highest for the 90\% load when the BFD is not used. This is originated by the fact that 80\% load is not heavy for the 10-sec segment size so that the quality change is low, while the quality is affected by the 90\% load that cause quality switches. Moreover, since 98\% load is the most influential for the quality, the quality cannot fluctuate so that the switch count is not higher than the case of 90\% load. Besides, considering the 1 second segment size, the count of switches between representations decrease for higher traffic loads since they cause worse video quality respectively.


\subsection{The Effect of BFD Intervals}

Our results showed that the impact of BFD is crucial in the case of congestion. In Figures \ref{49MbpsTimeDomain-10sec} and \ref{49MbpsTimeDomain-1sec}, the results show that using BFD fixes the poor outputs of each QoE parameters after the 150th second at which the traffic load starts to influence video streaming. Moreover, it is clearly observable that $T_i$ with 100ms outperformed $T_i$ with 1000ms regarding QoE parameters since its sensitivity is more delicate so that it detects the congestion earlier. Apart from the 49 Mbps shown in Figures \ref{49MbpsTimeDomain-10sec} and \ref{49MbpsTimeDomain-1sec}, this pattern is also the same for other traffic loads including 45 Mbps (90\% load) and 40 Mbps (80\% load). 

On the other hand, the effect of BFD is also noticeable in Figures \ref{ssim10sec} and  \ref{ssim1sec} considering the video quality based on the traffic load. For 10-sec segment size with 80\% traffic load, the effect of BFD interval is limited since the traffic could not cause congestion that induces to prevent BFD control packets considering their message interval,  $T_i$. However, for 1 second segment size with 80\% traffic load, $T_i$ with 100ms is affected by the traffic so that the quality is improved. However, considering the 90\% and 98\% traffic loads, $T_i$ with 1000ms is also affected by the congestion. Moreover, the number of quality switches between representations is reduced when we use BFD as shown in Figure \ref{switch10sec} and \ref{switch1sec}. The number of quality switches is the lowest for the $T_i$ with 100ms regarding 10-sec and 1 second segment sizes.

\section{Conclusion}

In this study, we investigated the fault tolerance in the SDN data plane considering the network-based parameters including the recovery time, packet loss, and throughput for QoS, and application-based parameters including the video quality value based on SSIM, and buffer level for QoE. Since network-based parameters for fault tolerance in the SDN data plane were studied before, we focused on two important issues that were not examined before. First, we focused on the dynamic protection considering quality of alternative paths for network-based parameters. Even though some studies mentioned the importance of the QoAPs, there is no study that implements this concept in the literature like our module, DPQoAP. Based on the results of our experiments, we clearly state that dynamic protection for fault tolerance is crucial as well as the recovery time. Second, we focused on the creation of the failure in a realistic manner which is not considered in the literature. Studies using an emulation environment such as Mininet run a special command to create link failures which also destroys the ports of the switches. Therefore, all of the other studies evaluated their results based on the Loss of Signal failure detection method which generally occurs for the port failures rather than the link failures. As a more realistic alternative, we used the Linux bridge, which is placed between the two switches and completely transparent to the controller, to evaluate a real-world case and obtain realistic results in Mininet. Hence, we created a more realistic comparison between the two cases. Our experiments showed that link failures carried out by the Mininet command are recovered within milliseconds while link failures created by shutting down one of the ports of the Linux bridge are recovered in seconds if BFD is not used. However, if BFD is used in the experiments using the Linux bridge and Mininet, the failure may be recovered faster based on the message transmit interval of the BFD protocol. 

We also studied the application-based parameters using DASH for fault tolerance. Since studies that focus on SDN data plane for fault tolerance considered only the network-based metrics, this is the first study that investigates how application and user experience are affected by the failure and the failure recovery mechanisms. To explore this, we applied the same scenarios and modules used for network-based parameters in a scenario with video streaming clients. To evaluate QoE, we used the video quality value based on SSIM and the buffer level in the client. The results of the experiments showed that the QoE appears to be not affected if the failure is created by the Mininet command since it is recovered within milliseconds. On the other hand, if the link failure is created by using the Linux bridge and BFD is not used, the video and therefore the QoE is affected for seconds. Moreover, we showed that if DPQoAP is not used for the recovery, the recovery time would not be important since the QoE significantly reduces because of the network conditions that affect the selected path.

On the other hand, we considered the congestion case applying the BFD protocol, which is originally designed to detect failures between network nodes, in order to detect the congestion on the path through which the video flows passing. We investigated the effect of the video segment size, traffic load, and BFD intervals on several QoE parameters that reflect the subjective opinion of the users. We used 1 second and 10 seconds long segment sizes; 100ms and 1000ms BFD intervals; 40 Mbps, 45 Mbps, 49 Mbps traffic loads with the capacity of 50 Mbps. Our results showed that QoE parameters of the video streaming with a large segment size is more stable than the small segment size for the congestion case. On the other hand, since the BFD interval with 100ms is more sensitive to the traffic load, it detects congestion earlier than the 1000ms interval so that the output of QoE parameters is better than the latter.

In the future, we plan to include reliability of the links into the fault tolerance problem. In this way, we believe that a more robust fault tolerant systems can be designed based on the probability of the link failures. Moreover, we plan to consider the number of DASH clients, the quality switch algorithm used in DASH and the percentage of rerouted flows to investigate their effect on the QoE parameters in case of congestion.

\section*{Acknowledgment}

This work is supported by the Turkish Directorate of Strategy and Budget under the TAM Project number DPT2007K120610, and the Galatasaray University Research Foundation under the Grant No. 18.401.003.

\bibliographystyle{elsarticle-harv} 

\bibliography{Software-Defined-Networks}

\begin{thebibliography}{47}
\expandafter\ifx\csname natexlab\endcsname\relax\def\natexlab#1{#1}\fi
\providecommand{\url}[1]{\texttt{#1}}
\providecommand{\href}[2]{#2}
\providecommand{\path}[1]{#1}
\providecommand{\DOIprefix}{doi:}
\providecommand{\ArXivprefix}{arXiv:}
\providecommand{\URLprefix}{URL: }
\providecommand{\Pubmedprefix}{pmid:}
\providecommand{\doi}[1]{\href{http://dx.doi.org/#1}{\path{#1}}}
\providecommand{\Pubmed}[1]{\href{pmid:#1}{\path{#1}}}
\providecommand{\bibinfo}[2]{#2}
\ifx\xfnm\relax \def\xfnm[#1]{\unskip,\space#1}\fi
\bibitem[{van Adrichem et~al.(2014)van Adrichem, van Asten and
  Kuipers}]{Adrichem2014}
\bibinfo{author}{van Adrichem, N.L.}, \bibinfo{author}{van Asten, B.J.},
  \bibinfo{author}{Kuipers, F.A.}, \bibinfo{year}{2014}.
\newblock \bibinfo{title}{{Fast Recovery in Software-Defined Networks}}, in:
  \bibinfo{booktitle}{2014 Third European Workshop on Software Defined
  Networks}, \bibinfo{publisher}{IEEE}. pp. \bibinfo{pages}{61--66}.
\newblock \URLprefix \url{http://ieeexplore.ieee.org/document/6984053/},
  \DOIprefix\doi{10.1109/EWSDN.2014.13}.
\bibitem[{Bagci et~al.(2017)Bagci, Sahin and Tekalp}]{bagci2017compete}
\bibinfo{author}{Bagci, K.T.}, \bibinfo{author}{Sahin, K.E.},
  \bibinfo{author}{Tekalp, A.M.}, \bibinfo{year}{2017}.
\newblock \bibinfo{title}{Compete or collaborate: Architectures for
  collaborative dash video over future networks}.
\newblock \bibinfo{journal}{IEEE Transactions on Multimedia}
  \bibinfo{volume}{19}, \bibinfo{pages}{2152--2165}.
\bibitem[{Bentaleb et~al.(2016)Bentaleb, Begen and
  Zimmermann}]{bentaleb2016sdndash}
\bibinfo{author}{Bentaleb, A.}, \bibinfo{author}{Begen, A.C.},
  \bibinfo{author}{Zimmermann, R.}, \bibinfo{year}{2016}.
\newblock \bibinfo{title}{Sdndash: Improving qoe of http adaptive streaming
  using software defined networking}, in: \bibinfo{booktitle}{Proceedings of
  the 2016 ACM on Multimedia Conference}, \bibinfo{organization}{ACM}. pp.
  \bibinfo{pages}{1296--1305}.
\bibitem[{Bentaleb et~al.(2017)Bentaleb, Begen, Zimmermann and
  Harous}]{bentaleb2017sdnhas}
\bibinfo{author}{Bentaleb, A.}, \bibinfo{author}{Begen, A.C.},
  \bibinfo{author}{Zimmermann, R.}, \bibinfo{author}{Harous, S.},
  \bibinfo{year}{2017}.
\newblock \bibinfo{title}{Sdnhas: An sdn-enabled architecture to optimize qoe
  in http adaptive streaming}.
\newblock \bibinfo{journal}{IEEE Transactions on Multimedia}
  \bibinfo{volume}{19}, \bibinfo{pages}{2136--2151}.
\bibitem[{Borokhovich et~al.(2014)Borokhovich, Schiff and
  Schmid}]{borokhovich2014provable}
\bibinfo{author}{Borokhovich, M.}, \bibinfo{author}{Schiff, L.},
  \bibinfo{author}{Schmid, S.}, \bibinfo{year}{2014}.
\newblock \bibinfo{title}{Provable data plane connectivity with local fast
  failover: Introducing openflow graph algorithms}, in:
  \bibinfo{booktitle}{Proceedings of the third workshop on Hot topics in
  software defined networking}, \bibinfo{organization}{ACM}. pp.
  \bibinfo{pages}{121--126}.
\bibitem[{Cascone et~al.(2017)Cascone, Sanvito, Pollini, Capone and
  Sans{\`o}}]{cascone2017fast}
\bibinfo{author}{Cascone, C.}, \bibinfo{author}{Sanvito, D.},
  \bibinfo{author}{Pollini, L.}, \bibinfo{author}{Capone, A.},
  \bibinfo{author}{Sans{\`o}, B.}, \bibinfo{year}{2017}.
\newblock \bibinfo{title}{Fast failure detection and recovery in sdn with
  stateful data plane}.
\newblock \bibinfo{journal}{International Journal of Network Management}
  \bibinfo{volume}{27}.
\bibitem[{Cheng et~al.(2017)Cheng, Zhang, Li, Yu, Lin and
  He}]{cheng2017congestion}
\bibinfo{author}{Cheng, Z.}, \bibinfo{author}{Zhang, X.}, \bibinfo{author}{Li,
  Y.}, \bibinfo{author}{Yu, S.}, \bibinfo{author}{Lin, R.},
  \bibinfo{author}{He, L.}, \bibinfo{year}{2017}.
\newblock \bibinfo{title}{Congestion-aware local reroute for fast failure
  recovery in software-defined networks}.
\newblock \bibinfo{journal}{IEEE/OSA Journal of Optical Communications and
  Networking} \bibinfo{volume}{9}, \bibinfo{pages}{934--944}.
\bibitem[{{Cisco Visual Networking}(2017)}]{index2017zettabyte}
\bibinfo{author}{{Cisco Visual Networking}}, \bibinfo{year}{2017}.
\newblock \bibinfo{title}{The zettabyte era: Trends and analysis}.
\newblock \bibinfo{journal}{Cisco white paper} \bibinfo{note}{Accessed on: June
  20, 2018. [Online]. Available: \url{https://bit.ly/2uBPyaa}}.
\bibitem[{{dash.js, 2019}()}]{dashjs}
\bibinfo{author}{{dash.js, 2019}}, .
\newblock \bibinfo{title}{DASH Industry Forum}.
\newblock \URLprefix \url{http://cdn.dashjs.org/latest/jsdoc/index.\\html}.
  \bibinfo{note}{{accessed at May 29, 2019}}.
\bibitem[{De~Oliveira et~al.(2014)De~Oliveira, Shinoda, Schweitzer and
  Prete}]{de2014using}
\bibinfo{author}{De~Oliveira, R.L.S.}, \bibinfo{author}{Shinoda, A.A.},
  \bibinfo{author}{Schweitzer, C.M.}, \bibinfo{author}{Prete, L.R.},
  \bibinfo{year}{2014}.
\newblock \bibinfo{title}{Using mininet for emulation and prototyping
  software-defined networks}, in: \bibinfo{booktitle}{Communications and
  Computing (COLCOM), 2014 IEEE Colombian Conference on},
  \bibinfo{organization}{IEEE}. pp. \bibinfo{pages}{1--6}.
\bibitem[{Desai and Nandagopal(2010)}]{desai2010coping}
\bibinfo{author}{Desai, M.}, \bibinfo{author}{Nandagopal, T.},
  \bibinfo{year}{2010}.
\newblock \bibinfo{title}{Coping with link failures in centralized control
  plane architectures}, in: \bibinfo{booktitle}{Communication Systems and
  Networks (COMSNETS), 2010 Second International Conference on},
  \bibinfo{organization}{IEEE}. pp. \bibinfo{pages}{1--10}.
\bibitem[{{Floodlight Controller, 2019}()}]{floodlight18}
\bibinfo{author}{{Floodlight Controller, 2019}}, .
\newblock \bibinfo{title}{Project Floodlight}.
\newblock \URLprefix \url{http://www.projectfloodlight.org\\/floodlight/}.
  \bibinfo{note}{{accessed at May 29, 2019}}.
\bibitem[{Fonseca and Mota(2017)}]{fonseca2017survey}
\bibinfo{author}{Fonseca, P.}, \bibinfo{author}{Mota, E.},
  \bibinfo{year}{2017}.
\newblock \bibinfo{title}{A survey on fault management in software-defined
  networks}.
\newblock \bibinfo{journal}{IEEE Communications Surveys \& Tutorials} .
\bibitem[{Georgopoulos et~al.(2013)Georgopoulos, Elkhatib, Broadbent, Mu and
  Race}]{georgopoulos2013towards}
\bibinfo{author}{Georgopoulos, P.}, \bibinfo{author}{Elkhatib, Y.},
  \bibinfo{author}{Broadbent, M.}, \bibinfo{author}{Mu, M.},
  \bibinfo{author}{Race, N.}, \bibinfo{year}{2013}.
\newblock \bibinfo{title}{Towards network-wide qoe fairness using
  openflow-assisted adaptive video streaming}, in:
  \bibinfo{booktitle}{Proceedings of the 2013 ACM SIGCOMM workshop on Future
  human-centric multimedia networking}, \bibinfo{organization}{ACM}. pp.
  \bibinfo{pages}{15--20}.
\bibitem[{Gozdecki et~al.(2003)Gozdecki, Jajszczyk and
  Stankiewicz}]{gozdecki2003quality}
\bibinfo{author}{Gozdecki, J.}, \bibinfo{author}{Jajszczyk, A.},
  \bibinfo{author}{Stankiewicz, R.}, \bibinfo{year}{2003}.
\newblock \bibinfo{title}{Quality of service terminology in ip networks}.
\newblock \bibinfo{journal}{IEEE Communications Magazine} \bibinfo{volume}{41},
  \bibinfo{pages}{153--159}.
\bibitem[{Gude et~al.(2008)Gude, Koponen, Pettit, Pfaff, Casado, McKeown and
  Shenker}]{gude2008nox}
\bibinfo{author}{Gude, N.}, \bibinfo{author}{Koponen, T.},
  \bibinfo{author}{Pettit, J.}, \bibinfo{author}{Pfaff, B.},
  \bibinfo{author}{Casado, M.}, \bibinfo{author}{McKeown, N.},
  \bibinfo{author}{Shenker, S.}, \bibinfo{year}{2008}.
\newblock \bibinfo{title}{Nox: towards an operating system for networks}.
\newblock \bibinfo{journal}{ACM SIGCOMM Computer Communication Review}
  \bibinfo{volume}{38}, \bibinfo{pages}{105--110}.
\bibitem[{{iPerf, 2019}()}]{iperf}
\bibinfo{author}{{iPerf, 2019}}, .
\newblock \bibinfo{title}{iPerf Traffic Generator}.
\newblock \URLprefix \url{https://iperf.fr}. \bibinfo{note}{{accessed at May
  29, 2019}}.
\bibitem[{Katz and Ward(2010)}]{rfc5880}
\bibinfo{author}{Katz, D.}, \bibinfo{author}{Ward, D.}, \bibinfo{year}{2010}.
\newblock \bibinfo{title}{{Bidirectional Forwarding Detection (BFD)}}.
\newblock \bibinfo{howpublished}{RFC 5880}.
\newblock \URLprefix \url{https://rfc-editor.org/rfc/rfc5880.txt},
  \DOIprefix\doi{10.17487/RFC5880}.
\bibitem[{Kempf et~al.(2012)Kempf, Bellagamba, Kern, Jocha, Tak{\'a}cs and
  Sk{\"o}ldstr{\"o}m}]{kempf2012scalable}
\bibinfo{author}{Kempf, J.}, \bibinfo{author}{Bellagamba, E.},
  \bibinfo{author}{Kern, A.}, \bibinfo{author}{Jocha, D.},
  \bibinfo{author}{Tak{\'a}cs, A.}, \bibinfo{author}{Sk{\"o}ldstr{\"o}m, P.},
  \bibinfo{year}{2012}.
\newblock \bibinfo{title}{Scalable fault management for openflow}, in:
  \bibinfo{booktitle}{Communications (ICC), 2012 IEEE international conference
  on}, \bibinfo{organization}{IEEE}. pp. \bibinfo{pages}{6606--6610}.
\bibitem[{Kim et~al.(2012)Kim, Schlansker, Santos, Tourrilhes, Turner and
  Feamster}]{kim2012coronet}
\bibinfo{author}{Kim, H.}, \bibinfo{author}{Schlansker, M.},
  \bibinfo{author}{Santos, J.R.}, \bibinfo{author}{Tourrilhes, J.},
  \bibinfo{author}{Turner, Y.}, \bibinfo{author}{Feamster, N.},
  \bibinfo{year}{2012}.
\newblock \bibinfo{title}{Coronet: Fault tolerance for software defined
  networks}, in: \bibinfo{booktitle}{Network Protocols (ICNP), 2012 20th IEEE
  International Conference on}, \bibinfo{organization}{IEEE}. pp.
  \bibinfo{pages}{1--2}.
\bibitem[{Kim et~al.(2016)Kim, Son, Talukder and Hong}]{kim2016congestion}
\bibinfo{author}{Kim, S.}, \bibinfo{author}{Son, J.},
  \bibinfo{author}{Talukder, A.}, \bibinfo{author}{Hong, C.S.},
  \bibinfo{year}{2016}.
\newblock \bibinfo{title}{Congestion prevention mechanism based on q-leaning
  for efficient routing in sdn}, in: \bibinfo{booktitle}{Information Networking
  (ICOIN), 2016 International Conference on}, \bibinfo{organization}{IEEE}. pp.
  \bibinfo{pages}{124--128}.
\bibitem[{Li et~al.(2014)Li, Hyun, Yoo, Baik and Hong}]{li2014scalable}
\bibinfo{author}{Li, J.}, \bibinfo{author}{Hyun, J.}, \bibinfo{author}{Yoo,
  J.H.}, \bibinfo{author}{Baik, S.}, \bibinfo{author}{Hong, J.W.K.},
  \bibinfo{year}{2014}.
\newblock \bibinfo{title}{Scalable failover method for data center networks
  using openflow}, in: \bibinfo{booktitle}{Network Operations and Management
  Symposium (NOMS), 2014 IEEE}, \bibinfo{organization}{IEEE}. pp.
  \bibinfo{pages}{1--6}.
\bibitem[{Lu and Zhu(2015)}]{lu2015sdn}
\bibinfo{author}{Lu, Y.}, \bibinfo{author}{Zhu, S.}, \bibinfo{year}{2015}.
\newblock \bibinfo{title}{Sdn-based tcp congestion control in data center
  networks}, in: \bibinfo{booktitle}{Computing and Communications Conference
  (IPCCC), 2015 IEEE 34th International Performance},
  \bibinfo{organization}{IEEE}. pp. \bibinfo{pages}{1--7}.
\bibitem[{Mkwawa et~al.(2016)Mkwawa, Barakabitze and Sun}]{mkwawa2016video}
\bibinfo{author}{Mkwawa, I.H.}, \bibinfo{author}{Barakabitze, A.A.},
  \bibinfo{author}{Sun, L.}, \bibinfo{year}{2016}.
\newblock \bibinfo{title}{Video quality management over the software defined
  networking}, in: \bibinfo{booktitle}{Multimedia (ISM), 2016 IEEE
  International Symposium on}, \bibinfo{organization}{IEEE}. pp.
  \bibinfo{pages}{559--564}.
\bibitem[{Nasimi et~al.(2018)Nasimi, Habibi, Han and Schotten}]{nasimi2018edge}
\bibinfo{author}{Nasimi, M.}, \bibinfo{author}{Habibi, M.A.},
  \bibinfo{author}{Han, B.}, \bibinfo{author}{Schotten, H.D.},
  \bibinfo{year}{2018}.
\newblock \bibinfo{title}{Edge-assisted congestion control mechanism for 5g
  network using software-defined networking}, in: \bibinfo{booktitle}{2018 15th
  International Symposium on Wireless Communication Systems (ISWCS)},
  \bibinfo{organization}{IEEE}. pp. \bibinfo{pages}{1--5}.
\bibitem[{Nguyen et~al.(2013)Nguyen, Minh and Yamada}]{nguyen2013software}
\bibinfo{author}{Nguyen, K.}, \bibinfo{author}{Minh, Q.T.},
  \bibinfo{author}{Yamada, S.}, \bibinfo{year}{2013}.
\newblock \bibinfo{title}{A software-defined networking approach for
  disaster-resilient wans}, in: \bibinfo{booktitle}{Computer Communications and
  Networks (ICCCN), 2013 22nd International Conference on},
  \bibinfo{organization}{IEEE}. pp. \bibinfo{pages}{1--5}.
\bibitem[{{Open vSwitch, 2019}()}]{openvswitch}
\bibinfo{author}{{Open vSwitch, 2019}}, .
\newblock \bibinfo{title}{Open Virtual Switch}.
\newblock \URLprefix \url{http://www.openvswitch.org}. \bibinfo{note}{{accessed
  at May 29, 2019}}.
\bibitem[{Oyman and Singh(2012)}]{oyman2012quality}
\bibinfo{author}{Oyman, O.}, \bibinfo{author}{Singh, S.}, \bibinfo{year}{2012}.
\newblock \bibinfo{title}{Quality of experience for http adaptive streaming
  services}.
\newblock \bibinfo{journal}{IEEE Communications Magazine} \bibinfo{volume}{50}.
\bibitem[{Petroulakis et~al.(2017)Petroulakis, Spanoudakis and
  Askoxylakis}]{petroulakis2017fault}
\bibinfo{author}{Petroulakis, N.E.}, \bibinfo{author}{Spanoudakis, G.},
  \bibinfo{author}{Askoxylakis, I.G.}, \bibinfo{year}{2017}.
\newblock \bibinfo{title}{Fault tolerance using an sdn pattern framework}, in:
  \bibinfo{booktitle}{GLOBECOM 2017-2017 IEEE Global Communications
  Conference}, \bibinfo{organization}{IEEE}. pp. \bibinfo{pages}{1--6}.
\bibitem[{Pfeiffenberger et~al.(2015)Pfeiffenberger, Du, Arruda and
  Anzaloni}]{pfeiffenberger2015reliable}
\bibinfo{author}{Pfeiffenberger, T.}, \bibinfo{author}{Du, J.L.},
  \bibinfo{author}{Arruda, P.B.}, \bibinfo{author}{Anzaloni, A.},
  \bibinfo{year}{2015}.
\newblock \bibinfo{title}{Reliable and flexible communications for power
  systems: Fault-tolerant multicast with sdn/openflow}, in:
  \bibinfo{booktitle}{New Technologies, Mobility and Security (NTMS), 2015 7th
  International Conference on}, \bibinfo{organization}{IEEE}. pp.
  \bibinfo{pages}{1--6}.
\bibitem[{Ramos et~al.(2013a)Ramos, Martinello and Rothenberg}]{ramos2013data}
\bibinfo{author}{Ramos, R.M.}, \bibinfo{author}{Martinello, M.},
  \bibinfo{author}{Rothenberg, C.E.}, \bibinfo{year}{2013}a.
\newblock \bibinfo{title}{Data center fault-tolerant routing and forwarding: An
  approach based on encoded paths}, in: \bibinfo{booktitle}{Dependable
  Computing (LADC), 2013 Sixth Latin-American Symposium on},
  \bibinfo{organization}{IEEE}. pp. \bibinfo{pages}{104--113}.
\bibitem[{Ramos et~al.(2013b)Ramos, Martinello and
  Rothenberg}]{ramos2013slickflow}
\bibinfo{author}{Ramos, R.M.}, \bibinfo{author}{Martinello, M.},
  \bibinfo{author}{Rothenberg, C.E.}, \bibinfo{year}{2013}b.
\newblock \bibinfo{title}{Slickflow: Resilient source routing in data center
  networks unlocked by openflow}, in: \bibinfo{booktitle}{Local Computer
  Networks (LCN), 2013 IEEE 38th Conference on}, \bibinfo{organization}{IEEE}.
  pp. \bibinfo{pages}{606--613}.
\bibitem[{Reitblatt et~al.(2013)Reitblatt, Canini, Guha and
  Foster}]{reitblatt2013fattire}
\bibinfo{author}{Reitblatt, M.}, \bibinfo{author}{Canini, M.},
  \bibinfo{author}{Guha, A.}, \bibinfo{author}{Foster, N.},
  \bibinfo{year}{2013}.
\newblock \bibinfo{title}{Fattire: Declarative fault tolerance for
  software-defined networks}, in: \bibinfo{booktitle}{Proceedings of the second
  ACM SIGCOMM workshop on Hot topics in software defined networking},
  \bibinfo{organization}{ACM}. pp. \bibinfo{pages}{109--114}.
\bibitem[{Seufert et~al.(2015)Seufert, Egger, Slanina, Zinner, Hossfeld and
  Tran-Gia}]{seufert2015survey}
\bibinfo{author}{Seufert, M.}, \bibinfo{author}{Egger, S.},
  \bibinfo{author}{Slanina, M.}, \bibinfo{author}{Zinner, T.},
  \bibinfo{author}{Hossfeld, T.}, \bibinfo{author}{Tran-Gia, P.},
  \bibinfo{year}{2015}.
\newblock \bibinfo{title}{A survey on quality of experience of http adaptive
  streaming}.
\newblock \bibinfo{journal}{IEEE Communications Surveys \& Tutorials}
  \bibinfo{volume}{17}, \bibinfo{pages}{469--492}.
\bibitem[{Sharma et~al.(2011)Sharma, Staessens, Colle, Pickavet and
  Demeester}]{sharma2011enabling}
\bibinfo{author}{Sharma, S.}, \bibinfo{author}{Staessens, D.},
  \bibinfo{author}{Colle, D.}, \bibinfo{author}{Pickavet, M.},
  \bibinfo{author}{Demeester, P.}, \bibinfo{year}{2011}.
\newblock \bibinfo{title}{Enabling fast failure recovery in openflow networks},
  in: \bibinfo{booktitle}{Design of Reliable Communication Networks (DRCN),
  2011 8th International Workshop on the}, \bibinfo{organization}{IEEE}. pp.
  \bibinfo{pages}{164--171}.
\bibitem[{Sharma et~al.(2013a)Sharma, Staessens, Colle, Pickavet and
  Demeester}]{sharma2013fast}
\bibinfo{author}{Sharma, S.}, \bibinfo{author}{Staessens, D.},
  \bibinfo{author}{Colle, D.}, \bibinfo{author}{Pickavet, M.},
  \bibinfo{author}{Demeester, P.}, \bibinfo{year}{2013}a.
\newblock \bibinfo{title}{Fast failure recovery for in-band openflow networks},
  in: \bibinfo{booktitle}{Design of reliable communication networks (drcn),
  2013 9th international conference on the}, \bibinfo{organization}{IEEE}. pp.
  \bibinfo{pages}{52--59}.
\bibitem[{Sharma et~al.(2013b)Sharma, Staessens, Colle, Pickavet and
  Demeester}]{sharma2013openflow}
\bibinfo{author}{Sharma, S.}, \bibinfo{author}{Staessens, D.},
  \bibinfo{author}{Colle, D.}, \bibinfo{author}{Pickavet, M.},
  \bibinfo{author}{Demeester, P.}, \bibinfo{year}{2013}b.
\newblock \bibinfo{title}{Openflow: Meeting carrier-grade recovery
  requirements}.
\newblock \bibinfo{journal}{Computer Communications} \bibinfo{volume}{36},
  \bibinfo{pages}{656--665}.
\bibitem[{Song et~al.(2017)Song, Park, Choi, Choi and Zhu}]{song2017control}
\bibinfo{author}{Song, S.}, \bibinfo{author}{Park, H.}, \bibinfo{author}{Choi,
  B.Y.}, \bibinfo{author}{Choi, T.}, \bibinfo{author}{Zhu, H.},
  \bibinfo{year}{2017}.
\newblock \bibinfo{title}{Control path management framework for enhancing
  software-defined network (sdn) reliability}.
\newblock \bibinfo{journal}{IEEE Transactions on Network and Service
  Management} \bibinfo{volume}{14}, \bibinfo{pages}{302--316}.
\bibitem[{Stockhammer(2011)}]{stockhammer2011dynamic}
\bibinfo{author}{Stockhammer, T.}, \bibinfo{year}{2011}.
\newblock \bibinfo{title}{Dynamic adaptive streaming over http--: standards and
  design principles}, in: \bibinfo{booktitle}{Proceedings of the second annual
  ACM conference on Multimedia systems}, \bibinfo{organization}{ACM}. pp.
  \bibinfo{pages}{133--144}.
\bibitem[{Thorat et~al.(2017)Thorat, Raza, Kim and Choo}]{thorat2017rapid}
\bibinfo{author}{Thorat, P.}, \bibinfo{author}{Raza, S.}, \bibinfo{author}{Kim,
  D.S.}, \bibinfo{author}{Choo, H.}, \bibinfo{year}{2017}.
\newblock \bibinfo{title}{Rapid recovery from link failures in software-defined
  networks}.
\newblock \bibinfo{journal}{Journal of Communications and Networks}
  \bibinfo{volume}{19}, \bibinfo{pages}{648--665}.
\bibitem[{Van~Adrichem et~al.(2014)Van~Adrichem, Van~Asten and
  Kuipers}]{van2014fast}
\bibinfo{author}{Van~Adrichem, N.L.}, \bibinfo{author}{Van~Asten, B.J.},
  \bibinfo{author}{Kuipers, F.A.}, \bibinfo{year}{2014}.
\newblock \bibinfo{title}{Fast recovery in software-defined networks}, in:
  \bibinfo{booktitle}{Software Defined Networks (EWSDN), 2014 Third European
  Workshop on}, \bibinfo{organization}{IEEE}. pp. \bibinfo{pages}{61--66}.
\bibitem[{Varis(2012)}]{varis2012anatomy}
\bibinfo{author}{Varis, N.}, \bibinfo{year}{2012}.
\newblock \bibinfo{title}{Anatomy of a linux bridge}, in:
  \bibinfo{booktitle}{Proceedings of Seminar on Network Protocols in Operating
  Systems}, p.~\bibinfo{pages}{58}.
\bibitem[{Wang et~al.(2004)Wang, Bovik, Sheikh and Simoncelli}]{wang2004image}
\bibinfo{author}{Wang, Z.}, \bibinfo{author}{Bovik, A.C.},
  \bibinfo{author}{Sheikh, H.R.}, \bibinfo{author}{Simoncelli, E.P.},
  \bibinfo{year}{2004}.
\newblock \bibinfo{title}{Image quality assessment: from error visibility to
  structural similarity}.
\newblock \bibinfo{journal}{IEEE Transactions on Image Processing}
  \bibinfo{volume}{13}, \bibinfo{pages}{600--612}.
\bibitem[{{Wireshark, 2019}()}]{wireshark}
\bibinfo{author}{{Wireshark, 2019}}, .
\newblock \bibinfo{title}{Wireshark Network Protocol Analyzer}.
\newblock \URLprefix \url{https://www.wireshark.org}. \bibinfo{note}{{accessed
  at May 29, 2019}}.
\bibitem[{Yu et~al.(2018)Yu, Li, Leng, Song, Bu, Chen, Yang, Zhang, Cheng and
  Xiao}]{yu2018fault}
\bibinfo{author}{Yu, Y.}, \bibinfo{author}{Li, X.}, \bibinfo{author}{Leng, X.},
  \bibinfo{author}{Song, L.}, \bibinfo{author}{Bu, K.}, \bibinfo{author}{Chen,
  Y.}, \bibinfo{author}{Yang, J.}, \bibinfo{author}{Zhang, L.},
  \bibinfo{author}{Cheng, K.}, \bibinfo{author}{Xiao, X.},
  \bibinfo{year}{2018}.
\newblock \bibinfo{title}{Fault management in software-defined networking: A
  survey}.
\newblock \bibinfo{journal}{IEEE Communications Surveys \& Tutorials} .
\bibitem[{Yuan et~al.(2018)Yuan, Jin, Zou, Yang and Yu}]{yuan2018practical}
\bibinfo{author}{Yuan, B.}, \bibinfo{author}{Jin, H.}, \bibinfo{author}{Zou,
  D.}, \bibinfo{author}{Yang, L.T.}, \bibinfo{author}{Yu, S.},
  \bibinfo{year}{2018}.
\newblock \bibinfo{title}{A practical byzantine-based approach for faulty
  switch tolerance in software-defined networks}.
\newblock \bibinfo{journal}{IEEE Transactions on Network and Service
  Management} \bibinfo{volume}{15}, \bibinfo{pages}{825--839}.
\bibitem[{Zabrovskiy et~al.(2016)Zabrovskiy, Kuzmin, Petrov and
  Fomichev}]{zabrovskiy2016emulation}
\bibinfo{author}{Zabrovskiy, A.}, \bibinfo{author}{Kuzmin, E.},
  \bibinfo{author}{Petrov, E.}, \bibinfo{author}{Fomichev, M.},
  \bibinfo{year}{2016}.
\newblock \bibinfo{title}{Emulation of dynamic adaptive streaming over http
  with mininet}, in: \bibinfo{booktitle}{Proceedings of the 18th Conference of
  Open Innovations Association FRUCT}, \bibinfo{organization}{FRUCT Oy}. pp.
  \bibinfo{pages}{391--396}.

\end{thebibliography}

\end{document}